\documentclass[12pt]{article} 
\usepackage[koi8-r]{inputenc}
\usepackage[english]{babel}
\usepackage{pazh}
\usepackage{graphicx,amssymb,amsmath}
\usepackage{natbib}
\usepackage{xspace}
\tightenlines

\newcommand{\Teff}{$T_{\rm eff}$}
\newcommand{\lgg}{log\,$g$}

\newcommand{\eps}[1]{\log\varepsilon_{\rm #1}}
\newcommand{\kms}{${\rm km s^{-1}}$}
\newcommand{\kH}{$S_{\!\rm H}$}    
\newcommand{\Eexc}{$E_{\rm exc}$}

\newcommand{\eu}[5]{\mbox{$#1\,^#2{\rm #3}^{#4}_{\rm #5}$}}

\sloppypar

\begin{document}
	
\baselineskip 21pt

\title{\bf Non-LTE abundances of nitrogen in the Sun \\ and reference A--F type stars}

\author{\bf \hspace{-1.3cm}\copyright\, 2024 \ \
L.~I.~Mashonkina\affilmark{1*,2}, T.~A.~Ryabchikova\affilmark{1} }

\affil{
{\it Institute of Astronomy, RAS, Pyatnitskaya st. 48, 119017 Moscow, Russia$^1$ \\
 Institute of Laser Physics, SB RAS, Ac. Lavrentieva ave. 13, Novosibirsk 630090, Russia$^2$}
}

	\vspace{2mm}
	
	\sloppypar 
	\vspace{2mm}
	
A new model atom of nitrogen was developed using the energy levels of \ion{N}{1} from laboratory measurements and also predicted in atomic structure calculations and the most up-to-date atomic data for computing radiative and collisional rates of the transitions. Solar abundance $\eps{\odot,N}$(1D NLTE) = 7.92$\pm$0.03 was determined from lines of \ion{N}{1} by applying the synthetic spectrum method with the plane-parallel (1D) MARCS model atmosphere and taking into account the departures from local thermodynamic equilibrium (non-LTE effects). Having implemented the 3D-corrections of Amarsi et al. (2020), we obtained for the Sun $\eps{\odot,N}$(NLTE+3D) = 7.88$\pm$0.03. 
Based on high spectral resolution spectra, non-LTE abundances of nitrogen were derived for 11 unevolved A--F type stars with reliable atmospheric parameters. Non-LTE leads to strengthened \ion{N}{1} lines, and the non-LTE effects grow with increasing effective temperature. For each star, non-LTE leads to smaller abundance error compared to the LTE case. For superficially normal A-type stars, non-LTE removes the enhancements relative to the solar nitrogen abundance obtained in the LTE case. A $\lambda$~Boo-type star HD~172167 (Vega) also has close-to-solar N abundance. The four Am stars reveal a scatter of the N abundances, from [N/H] = $-0.44$ to [N/H] = 0.39. The N abundances of the Sun and superficially normal A stars are consistent within 0.09~dex with the nitrogen abundance of the interstellar gas and the early B-type stars.

 \noindent
{\bf Key words:\/} stellar atmospheres, non-LTE line formation, abundances of nitrogen in stars

\vfill
\noindent\rule{8cm}{1pt}\\
{$^*$ e-mail $<$lima@inasan.ru$>$}

\section{Introduction}

Nitrogen is one of the three most abundant elements (C, N, O) after hydrogen and helium. It plays an important role in the physics of stars, and knowledge of its abundance in a particular star, or rather, the change compared to the initial abundance at the birth of the star, is the key to understanding the evolutionary status of the star and the physical processes occurring in it. One believes that the chemical composition of galactic matter changed very little after the formation of the Solar System, therefore, for most chemical elements, their meteoritic abundances serve as abundance standards for the present-day Galactic matter or as cosmic abundance standards. But nitrogen is a volatile element, and the cosmic abundance standard for N is determined by the Sun and young, unevolved stars.

In the solar spectrum, nitrogen is observed in a few, weak and blended lines of \ion{N}{1}, as well as in the molecular lines of NH and CN. For many years, the abundance $\eps{\odot,N} = 8.05\pm0.04$ obtained by \cite{cosmos89} from molecular lines using the semi-empirical solar model atmosphere of \cite{HM74} was considered as standard. Hereafter, we use the abundance scale where $\eps{H}$ = 12. With the same model atmosphere, but from the \ion{N}{1} lines, \cite{n1_sun1996} determined $\eps{\odot,N} = 8.05\pm0.09$ based on the non-local thermodynamic equilibrium (non-LTE, NLTE) line formation. In the early 2000s, 3-dimensional (3D) atmospheric models based on hydrodynamic calculations became widespread. Using the 3D model and non-LTE abundance corrections (differences between non-LTE and LTE abundances: $\Delta_{\rm NLTE} = \eps{NLTE} - \eps{LTE}$) for the \ion{N}{1} lines calculated with the 1D model, \cite{Caffau2009} derived $\eps{\odot,N}$ = 7.86. In self-consistent 3D and non-LTE calculations for \ion{N}{1}, \cite{Amarsi_n1} determined the lower abundance, $\eps{\odot,N}$ = 7.767, but a higher value of $\eps{\odot,N}$ = 7.89 (3D) was obtained from the NH and CN lines \citep{Amarsi_CN}. It is necessary to understand the reasons for such discrepancies between different authors and between atomic and molecular lines.

For A-type stars, non-LTE abundances from the \ion{N}{1} lines were derived by \cite{takeda1992,Lemke95,n1_sun1996,Przybilla_n1,korotin_n1} and in the later studies based on already developed non-LTE methods. The methods were tested on the star Vega (HD~172167) with reliable atmospheric parameters and a high-quality observed spectrum. In various works, non-LTE corrections for the same line differ by 0.2~dex or more. For example, for \ion{N}{1} 8683~\AA, \cite{n1_sun1996} and \cite{Przybilla_n1} give $\Delta_{\rm NLTE} = -0.53$~dex and $-0.26$~dex, respectively. \cite{takeda1992} and \cite{Lemke95} drew attention to the sensitivity of non-LTE results for \ion{N}{1} to abundances of carbon and metals in the atmospheric model, i.e. to the  continuum absorption coefficient in the ultraviolet (UV) range.

For the star 21~Peg (HD~209459, B9.5~V), \cite{Roman2023} determined $\eps{NLTE}$ = 7.53 from the \ion{N}{1} lines. This is much lower than the lowest derived solar abundance. For comparison, \cite{Nieva2012} found $\eps{NLTE} = 7.79\pm0.04$ from lines of \ion{N}{2} in early  B-type stars.

Conflicting results for the Sun and young A--B stars motivated the present study.
The following tasks were set.

1. Construction of a new model atom for nitrogen to be applied to both solar type and A-type stars. The model atom should be as complete as possible and provide a close collisional coupling of the \ion{N}{1} high-excitation levels to the ground state of the next stage ionization \ion{N}{2}.

2. Determination of abundances from the \ion{N}{1} lines in the solar spectrum. We already have experience in determining the solar elemental non-LTE abundances using a classical one-dimensional (1D) model of the solar atmosphere from the MARCS model grid \citep{2008A&A...486..951G}, and our results are in excellent agreement with self-consistent 3D and non-LTE calculations of \cite{Asplund2021}. Namely, \cite{c_sun2015} obtained the same abundance from  lines of \ion{C}{1} and molecular lines of C$_2$ and CH: $\eps{\odot,C} = 8.43\pm0.02$, which is by 0.03~dex less than that in \cite{Asplund2021}. With the same atmospheric model, \cite{o_sun2018} derived the non-LTE abundance from lines of \ion{O}{1}, $\eps{\odot,O} = 8.70\pm0.08$, which is consistent within 0.01~dex with that in \cite{Asplund2021}.

3. Determination of non-LTE abundances from the \ion{N}{1} lines in the selected stars of F, A and late B-type. This is an extension of our previous detailed studies of the chemical composition of these stars. The results of non-LTE abundance determinations for He, C, O, Na, Mg, Si, Ca, Ti, Sr, Zr, and Ba are summarized by Mashonkina et al. (2020) and Romanovskaya et al. (2023). \cite{zn2022} and \cite{mash_sc} determined the non-LTE abundances of Zn and Sc.

This article is structured as follows. New model atom of \ion{N}{1} is presented in Sect.~\ref{sect:method}. The developed method is used to determine the nitrogen non-LTE abundances for the solar atmosphere (Sect.~\ref{sect:sun}) and for a sample of A-F type stars with reliably determined atmospheric parameters (Sect.~\ref{sect:sample}). Conclusions are formulated in Sect~\ref{conclusions}.

\section{Method of non-LTE calculations for \ion{N}{1}}\label{sect:method}

\subsection{New model atom of nitrogen}

\begin{figure}  
	\centering
	\includegraphics[width=0.90\columnwidth,clip]{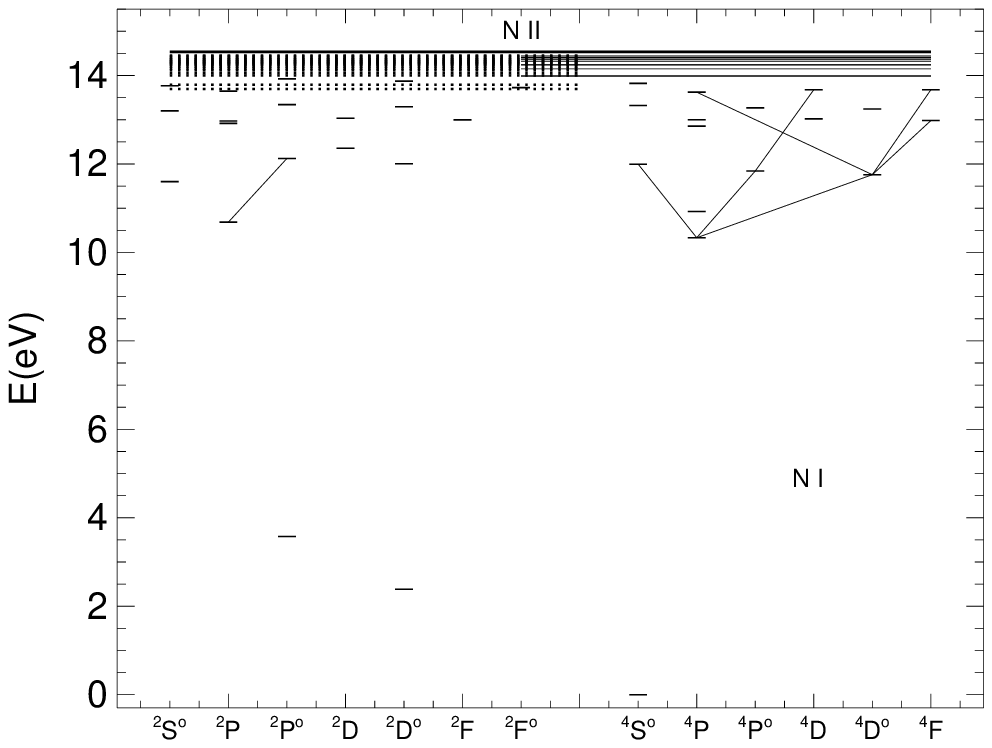}
	\caption{Energy levels in the nitrogen model atom and transitions, in which the \ion{N}{1} lines used to determine the stellar nitrogen abundances are formed. Solid and dotted horizontal segments indicate the energies of even and odd superlevels, respectively. The coordinates of the superlevels on the X-axis are assigned arbitrarily to minimize the overlap of even and odd levels.}
	\label{fig:atom_n1}
\end{figure}

{\bf Energy levels.} The atomic model is built using 321 energy levels of \ion{N}{1} from the NIST\footnote{https://www.nist.gov/pml/atomic-spectra-database} database (Kramida et al. 2019) and 503 levels predicted by Kurucz (2014) in the atomic structure calculations for \ion{N}{1}, but not detected (yet) in laboratory measurements. The excitation energies of the highest levels are less than the ionization energy of \ion{N}{1} by 0.08-0.26~eV, which is much smaller than the average kinetic energy of electrons at temperatures up to 20\,000~K. This ensures effective collisional coupling of \ion{N}{1} with the ground state of \ion{N}{2}. Multiplet fine structure was neglected. Highly excited levels with low energy separation ($<$250 cm$^{-1}$) and of the same parity were combined into superlevels. They are formed predominantly from levels predicted in calculations of the \ion{N}{1} atomic structure. The average energy of the combined superlevel was calculated taking into account the statistical weights of individual levels. The final atomic model includes 59 levels of \ion{N}{1}. Singly ionized nitrogen is represented only by its ground state. Excited levels increase the partition function of \ion{N}{2} by about 5\%\ at $T$ = 10000~K. Test calculations have shown that taking into account excited levels of \ion{N}{2} has a negligible effect on the magnitude of deviations from LTE for levels of \ion{N}{1}. The model atom is  shown in Fig.~\ref{fig:atom_n1}.

{\bf Radiative rates.} $gf$-values for 46\,109 bound-bound (b-b) transitions between levels used in constructing the atomic model are taken from calculations by \cite{K09}. After combining the levels, we got 743 b-b transitions with the oscillator strength $f > 10^{-6}$. For transitions from the ground state \eu{2p^3}{4}{S}{\circ}{} and low-excitation levels \eu{2p^3}{2}{D}{\circ}{}, \eu{2p^3}{2}{P}{\circ}{}, in which UV pumping of the upper levels can occur, radiative rates are calculated using the Voigt absorption profile (in total, for 16 transitions). For other transitions, the absorption profile is the Doppler one.

For 33 levels with the ionization thresholds $\lambda_{\rm thr} \le 18723$~\AA, the photoionization cross-sections calculated in the Opacity Project \citep{seaton87} are taken from the TOPbase database \citep{topbase}. For other levels, the cross-sections are calculated in the hydrogenic approximation using the effective principal quantum number instead of the principal quantum number.

{\bf Collisional rates.} For electron-impact excitation, we use the data of \citet{Wang_n1}, calculated with one of the variants of the R-matrix method (B-spline R-matrix with pseudo states). They are available for 276 transitions between levels with excitation energy \Eexc\ $\le$ 13.34~eV. For the remaining b-b transitions, we apply formula of van Regemorter (1962), if the transition is allowed, and assume the effective collision strength $\Omega$ = 1 for the  forbidden transitions. Electron-impact ionization rate is calculated using the \cite{seaton62} formula witn an adopted photoionization cross-section at the thresholds.

In the atmospheres of late-type stars, the number density of free electrons is much lower than that of neutral hydrogen atoms, so the excitation of levels and the formation of ions can be produced by collisions with not only electrons, but also hydrogen atoms. Rate coefficients for inelastic processes in collisions with \ion{H}{1} are taken from calculations of \citet{Amarsi_n1_hyd} for 21 transitions N$^0$ + H$^0$ $\leftrightarrow$ N$^+$ + H$^-$, associated with the S, P, D terms of \ion{N}{1}, and for 206 b-b transitions.

\begin{figure}  
	\hspace{-5mm}
	\vspace{-36mm}
	\centering
	\includegraphics[width=0.50\columnwidth,clip]{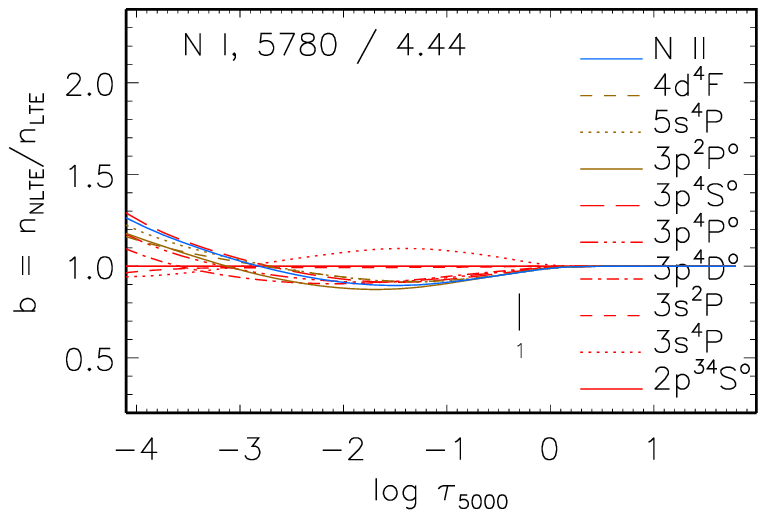}
	\includegraphics[width=0.50\columnwidth,clip]{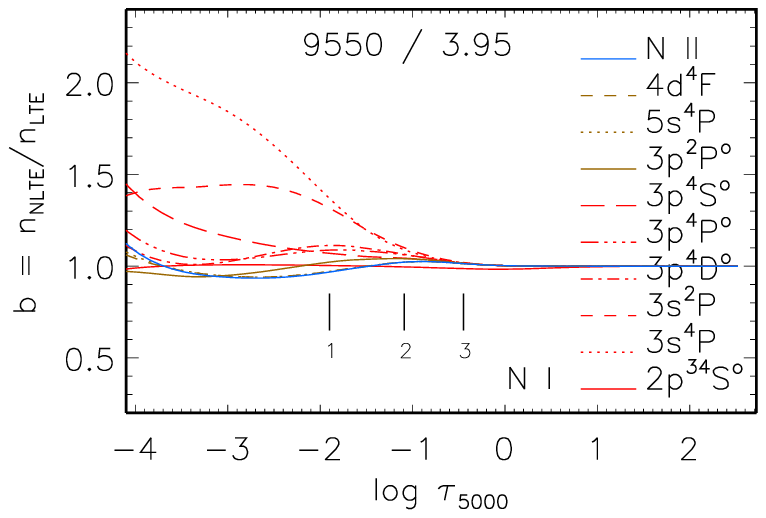}

	\vspace{36mm}
	\includegraphics[width=0.50\columnwidth,clip]{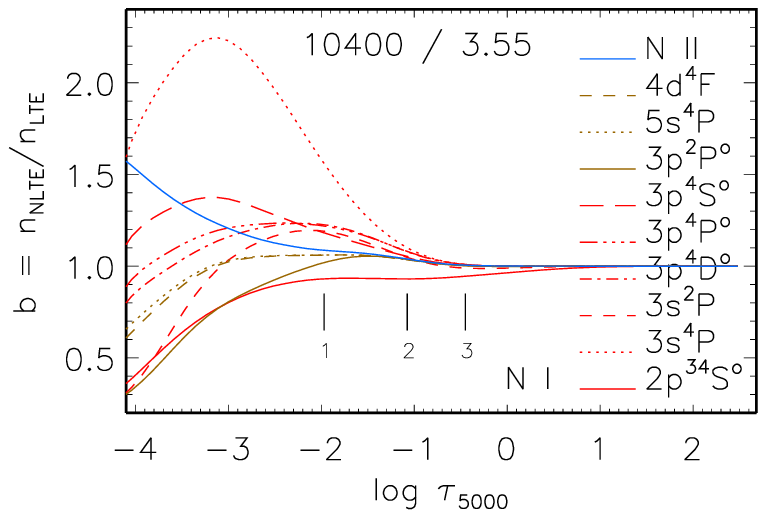}

	\caption{b-factors of selected levels of \ion{N}{1} as a function of optical depth at $\lambda$ = 5000~\AA\ in the atmospheric models 5780/4.44 (\Teff\ = 5780~K, \lgg\ = 4.44), 9550/3.95, and 10400/3.55. In each panel, tick marks indicate the locations of line center optical depth unity for the following lines: \ion{N}{1} 8683 (1), 8629 (2) and 6482~\AA\ (3). For the solar atmosphere, the formation depth is indicated only for the strongest line. }
	\label{fig:bf}
\end{figure}

\subsection{Statistical equilibrium of nitrogen depending on effective temperature}

To solve the system of statistical equilibrium (SE) and radiative transfer equations in a given atmospheric model, we use a modified code {\sc detail} \citep{giddings81,butler84,Przybilla2011}. We note that the line opacity is calculated directly, line-by-line using the list of lines by \cite{K1993}, which includes about 40 million lines.

Figure~\ref{fig:bf} shows b-factors, b = $n_{\rm NLTE} /n_{\rm LTE}$, for selected levels of \ion{N}{1}, which are important for understanding the mechanisms of departures from LTE in model atmospheres with different effective temperatures (\Teff). Here, $n_{\rm NLTE}$ and $n_{\rm LTE}$ are level populations, obtained by solving the SE equations (non-LTE) and using the Boltzmann-Saha formulas (LTE). Throughout this work we use plane-parallel (1D) model atmospheres. For the Sun, the model with \Teff\ = 5780~ K and surface gravity \lgg\ = 4.44 (5780/4.44) is taken from MARCS database\footnote{http://marcs.astro.uu.se} (Gustafsson et al.2008). The models 9550/3.95 and 10400/3.55 were calculated using the LL{\sc models} code \citep{llmodels}.

In an atmosphere with \Teff\ = 5780~K, neutral atoms dominate in the nitrogen abundance, therefore the ground and two lowest excited levels keep thermodynamic equilibrium populations and reveal b = 1 throughout the atmosphere. The lower level \eu{3s}{4}{P}{}{} of the transitions, in which the observed lines are formed (see Table~\ref{tab:sun}), is overpopulated (b $>$ 1) above log~$\tau_{5000}$ = 0 due to radiative pumping from the ground state in the wings of the \ion{N}{1} 1199.55, 1200.22, and 1200.71~\AA\ lines. The upper levels, e.g. \eu{3p}{4}{S}{\circ}{}, are depopulated (b $<$ 1) in spontaneous transitions to the lower states. As a result, lines of \ion{N}{1} are enhanced compared with the LTE case. But the departures from LTE are weak.

In the 9550/3.95 model atmosphere, \ion{N}{1} remains the dominant ionization stage, but radiative processes are strengthened compared to collisional ones, and the departures from LTE are stronger than in the case of the solar atmosphere. Not only the lower, but also the upper levels of the transitions, in which the observed lines are formed, are overpopulated, but to a lesser extent than the lower ones. Therefore, non-LTE leads to increased absorption in the lines.

In the highest temperature model (10400/3.55), neutral and singly ionized nitrogen have comparable number densities. For the low-excitation levels \eu{2p^3}{2}{P}{\circ}{} ($\lambda_{\rm thr} \simeq$ 1130~\AA) and \eu{2p^3}{2}{D}{\circ}{} ($\lambda_{\rm thr} \simeq$ 1020~\AA), the intensity of the ionizing radiation exceeds the local value of the Planck function  resulting in overionization of these levels. Due to their close coupling to the ground state, the latter is also emptied (b $<$ 1). Radiative pumping from the ground state leads to enhanced population of the high-excitation levels in such a way that, in each of the transitions associated with the observed lines, b$_{\rm low} >$ b$_{\rm up}$ (the lower and upper levels of the corresponding transitions are indicated in Table~\ref{tab:21peg}), and the lines are strengthened compared to the LTE case.

\begin{table}
	\centering
	\renewcommand{\arraystretch}{1.0}
	\renewcommand{\tabcolsep}{7pt}
	\caption{LTE and NLTE abundances derived from nitrogen lines in the solar spectrum.}
	\vspace{3mm}
	\label{tab:sun}
	\begin{tabular}{lcrcccc}
		\hline \noalign{\smallskip}
	\multicolumn{1}{c}{$\lambda$} & \Eexc & log $gf$ & transition  & \multicolumn{3}{c}{$\eps{}$}  \\
		\cline{5-7}
	\multicolumn{1}{c}{[\AA]}     &  [eV] &         &          & LTE & NLTE &  + 3D$^1$ \\
		\noalign{\smallskip} \hline \noalign{\smallskip}
\ion{N}{1}   7442.28 &  10.33 & --0.401 & \eu{3s}{4}{P}{}{3/2} -- \eu{3p}{4}{S}{\circ}{3/2} & 7.93 & 7.91 & 7.88 \\
\ion{N}{1}   8216.34 &  10.34 &  0.138 & \eu{3s}{4}{P}{}{5/2} -- \eu{3p}{4}{P}{\circ}{5/2}  & 8.00 & 7.95 & 7.92 \\
\ion{N}{1}   8629.24 &  10.69 &  0.077 & \eu{3s}{2}{P}{}{3/2} -- \eu{3p}{2}{P}{\circ}{3/2}  & 7.88 & 7.87 & 7.83 \\
\ion{N}{1}   8683.40 &  10.33 &  0.105 & \eu{3s}{4}{P}{}{3/2} -- \eu{3p}{4}{D}{\circ}{5/2}  & 7.95 & 7.90 & 7.87 \\
\ion{N}{1}  10108.89 &  11.75 &  0.443 & \eu{3p}{4}{D}{\circ}{3/2} -- \eu{3d}{4}{F}{}{5/2}  & 7.96 & 7.95 & 7.90 \\
\cline{5-7}
Mean	             &        &        &                                                    & 7.94 & 7.92 & 7.88 \\
 $\sigma$            &        &        &                                                    & 0.04 & 0.03 & 0.03 \\
\noalign{\smallskip}\hline \noalign{\smallskip}
CN lines            &        &        &                              & \multicolumn{2}{l}{7.97$\pm$0.04$^2$} & 7.91 \\
		\noalign{\smallskip}\hline \noalign{\smallskip}
\end{tabular}

 $^1$ after implementing the 3D NLTE corrections from \cite{Amarsi_n1} for lines of \ion{N}{1} and 3D corrections from \cite{Amarsi_CN} for molecular CN lines. $^2$ from \cite{Ryabchikova_CN}.
\end{table}

\section{Abundance of nitrogen in the solar atmosphere}\label{sect:sun}

\cite{Ryabchikova_CN} simultaneously determined the carbon and nitrogen abundances of the solar atmosphere, $\eps{C}$ = 8.44 and $\eps{N}$(CN) = 7.97, from molecular C$_2$ and CN lines by fitting the synthetic spectra to the observed one in 21 intervals in the range 5100-5200~\AA\ and 12 intervals in the range 7930-8100~\AA. According to \citet{Amarsi_CN}, the 3D correction to the N abundance derived from CN lines in the solar spectrum amounts to $-0.06$~dex, on average. Taking into account this 3D correction, \cite{Ryabchikova_CN} deduce $\eps{N}$(CN,3D) = 7.91$\pm$0.04.

\subsection{Analysis of the \ion{N}{1} lines in the solar spectrum}

We analyze the spectrum of the Sun as a star using the Atlas of Kurucz et al. (1984). Spectral resolving power is R = $\lambda/\Delta\lambda \simeq$ 520\,000. As in our previous studies, we use the canonical parameters of the solar atmosphere: \Teff\ = 5780~K, \lgg\ = 4.44, microturbulence velocity $\xi_{t}$ = 0.9~\kms\ and the classical atmospheric model from the MARCS database.

Abundances from the \ion{N}{1} lines were determined by the synthetic spectrum method, that is, by automatically fitting the theoretical spectrum to the observed one. The synthetic spectrum was calculated with the code \textsc{synthV\_NLTE} (Tymbal et al., 2019), integrated into the BinMag visualization program \citep{Kochukhov_binmag}. Using b-factors from the code {\sc detail}, \textsc{synthV\_NLTE} calculates lines of \ion{N}{1} taking into account the non-LTE effects, while lines of other elements are computed under the LTE assumption. Line atomic parameters for calculating the synthetic spectrum are taken from the current version of the database {\sc VALD3} \citep[Vienna Atomic Line Database,][]{vald_hfs}. For the \ion{N}{1} and CN molecular lines, we adopted $gf$-values from calculations by \citet{n1_nist} and \citet{CN}, respectively. In our calculations, we used a
fixed value of the solar carbon abundance, $\eps{C}$ = 8.43, as derived by \citet{c_sun2015}.

In the solar spectrum, there are no unblended (pure) lines of \ion{N}{1}. Five lines, which are most suitable for determining N abundance, are listed in Table~\ref{tab:sun}. Four of the five lines are blended with the CN molecular lines, and \ion{N}{1} 8216.336~\AA\ is blended with a weak \ion{Cr}{1} line. In addition, the \ion{N}{1} 8629.2 and 8683.4~\AA\ lines lie in the wings of the strong \ion{Ca}{2} 8662.14~\AA\ line, and the absorption in this line must be taken into account when fitting the synthetic spectrum to the observed one. In the same way, it is necessary to take into account the influence of nearby lines of other elements, primarily elements of the iron group, which can influence the position of the continuum near the \ion{N}{1} lines.

\begin{figure}  
	\centering
	\includegraphics[width=0.7\columnwidth,clip]{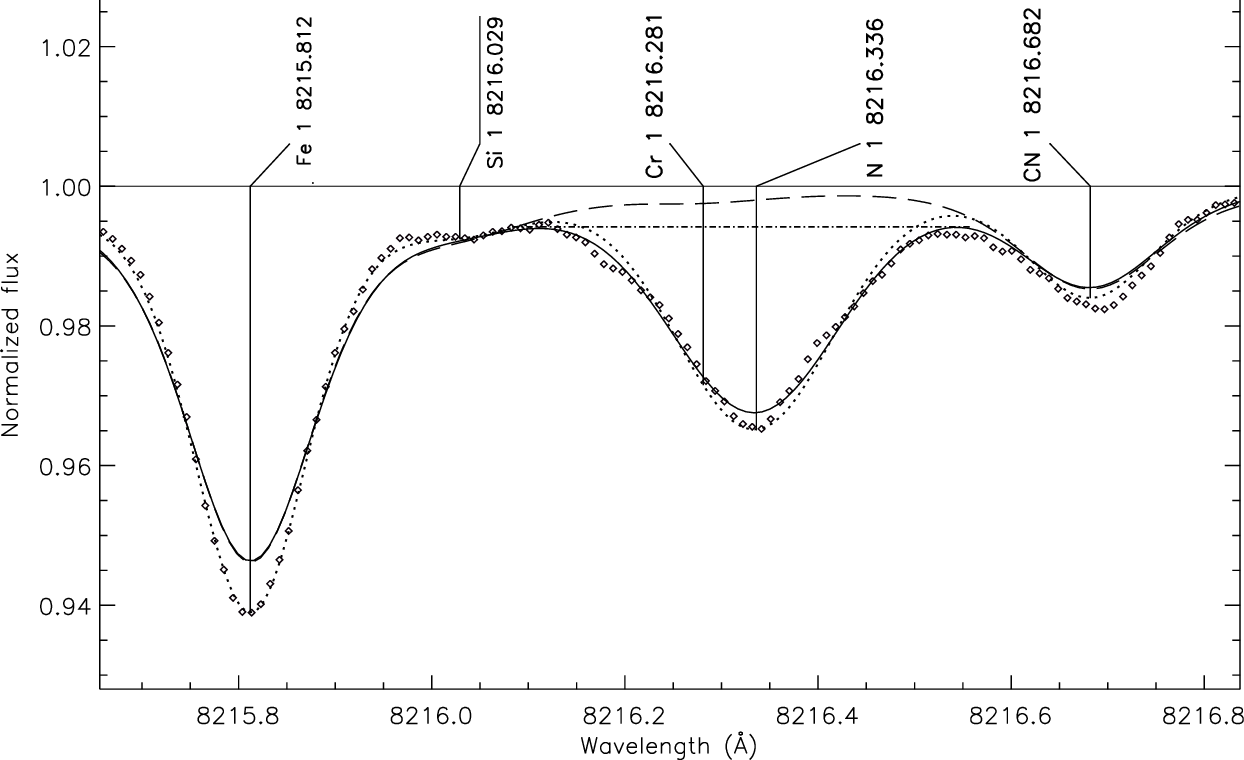}

	\vspace{4mm}
	\includegraphics[width=0.7\columnwidth,clip]{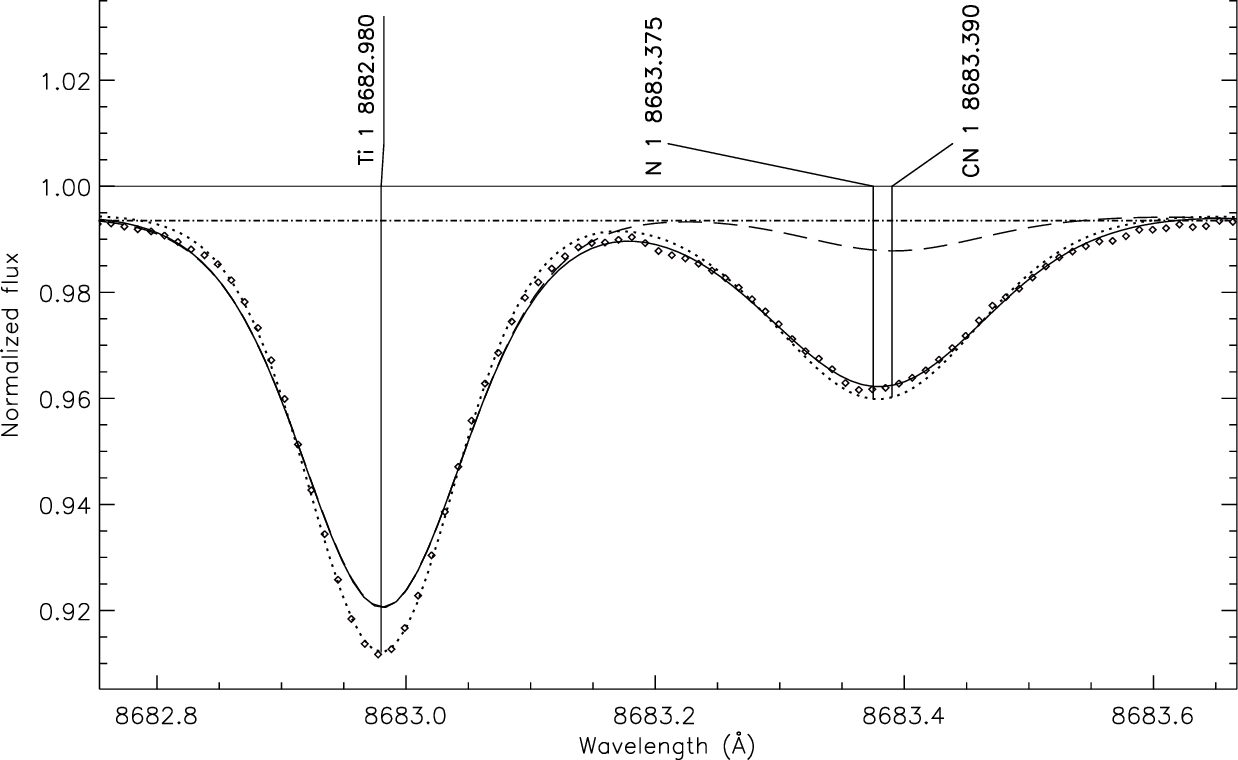}
	\caption{\ion{N}{1} 8216.336 and 8683.40~\AA\ lines in the solar flux spectrum (Kurucz et al. 1984, diamonds) compared to synthetic spectra, where the non-LTE effects are taken into account for the \ion{N}{1} lines. In each panel: the dotted curve is the best fit to the entire region (8215.6 -- 8216.8~\AA\ and 8682.7 -- 8683.7~\AA); solid curve is the best fit to the \ion{N}{1} line; dashed curve is the synthetic spectrum without the contribution of the \ion{N}{1} line; dash-dotted straight line shows the local continuum level used in our test calculations, which led to a decrease in N abundance, by 0.12~dex for \ion{N}{1} 8216~\AA\ and 0.05~dex for \ion{N}{1} 8683~\AA. See text for more details.}
	\label{fig:8216}
\end{figure}

Let us consider the fitting procedure separately for each line.

{\bf The \ion{N}{1} 8216.336~\AA\ line} is the most suitable because it is not blended by CN molecular lines. However, this line lies in a region with strong telluric lines, making it difficult to determine the continuum level. The position of the continuum was refined by comparing the solar spectrum from Kurucz et al. (1984) with the solar spectrum corrected by \citet{Sun_2011} for telluric lines. With the continuum level defined over 8215.6 -- 8216.8~\AA, the entire region was first fitted, and then separately profile of \ion{N}{1} 8216.336~\AA, as shown in Fig.~\ref{fig:8216}. The nitrogen abundance turned out to be the same, $\eps{N}$ = 7.95 (non-LTE), in both cases. Ignoring weak \ion{Cr}{1} 8216.281~\AA\ line leads to an increase in the nitrogen abundance, by no more than 0.01~dex. As can be seen in Fig.~\ref{fig:8216}, reproducing the profile shape of the nitrogen line requires a higher macroturbulent velocity than for the stronger \ion{Fe}{1} 8215.8~\AA\ line. This is a well-known fact. \cite{gray77} also showed that the profile shapes of strong and weak \ion{Fe}{1} lines in the solar spectrum are fitted with macroturbulent velocities of 3.1 and 3.8~\kms, respectively.

In order to understand how \citet[][Table~3, 1D non-LTE]{Amarsi_n1} obtained a low abundance of $\eps{N}$ = 7.796 for the same line and with the same atmospheric model as ours, we carried out test calculations by determining the continuum level along the edges of the \ion{N}{1} 8216~\AA\ line, i.e. using the local continuum, and derived $\eps{N}$ = 7.83, which is close to the N abundance published by \citet{Amarsi_n1}. In their abundance determinations, \citet{Amarsi_n1} used the average equivalent width $EW$ = 7.7~m\AA, measured in the two solar disk-center intensity spectra of \citet{Delbouille1973} and \citet{Neckel84}. \cite{Amarsi_n1} note that the difference in $EW$(\ion{N}{1} 8216~\AA) between the two spectra is less than 5\%, but \citet{sun1990} measured $EW$ = 8.7~m\AA\ for the same line in the spectrum of \cite{Delbouille1973}, which exceeds the value used by \cite{Amarsi_n1}, by 13\%. With such an overlap of lines of different elements, as in the region 8215.6 -- 8216.8~\AA, using $EW$ in abundance determinations is unjustified.

{\bf The \ion{N}{1} 8683.40~\AA\ line} is located in the wings of strong \ion{Ca}{2} 8662~\AA\ and \ion{H}{1} 8665~\AA\ lines. With the continuum level determined taking into account absorption in the \ion{Ca}{2} and \ion{H}{1} lines, we first fitted the spectral range 8682.7 -- 8683.7~\AA, and then separately profile of the \ion{N}{1} line (Fig.~\ref{fig:8216}). The obtained nitrogen abundance is indicated in Table~\ref{tab:sun}. Note that we had to reduce the wavelength of the \ion{N}{1} line by 0.025~\AA\ to match the wavelength  differences between the \ion{Fe}{1} 8674.746 and 8688.623~\AA\ lines, which have reliably measured wavelengths, and the \ion{N}{1} 8683.40~\AA\ line in the observed and synthetic spectra.
As in the case of \ion{N}{1} 8216~\AA, we also determined the nitrogen abundance using a local continuum, i.e. without taking into account absorption in lines of \ion{Ca}{2} and \ion{H}{1}, and obtained the lower abundance, by 0.05~dex.

\begin{figure}  
	\centering
	\includegraphics[width=0.7\columnwidth,clip]{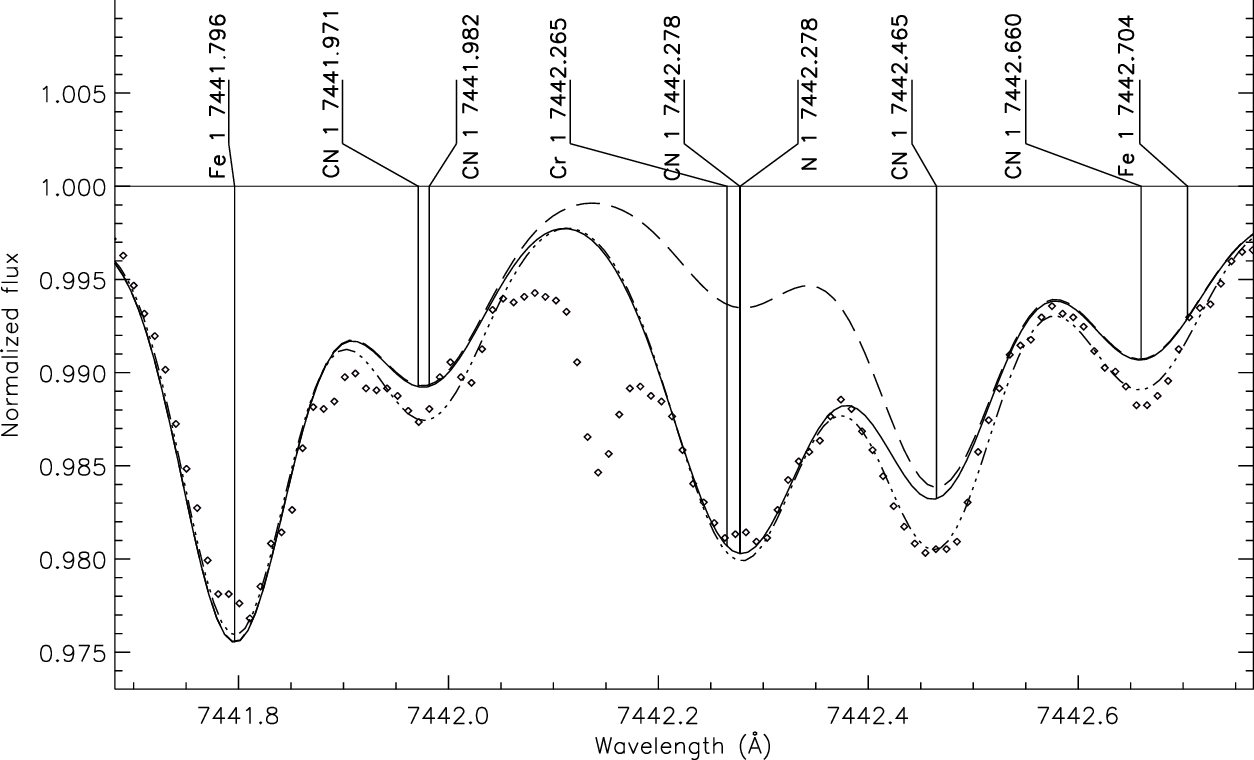}

	\vspace{4mm}
	\includegraphics[width=0.7\columnwidth,clip]{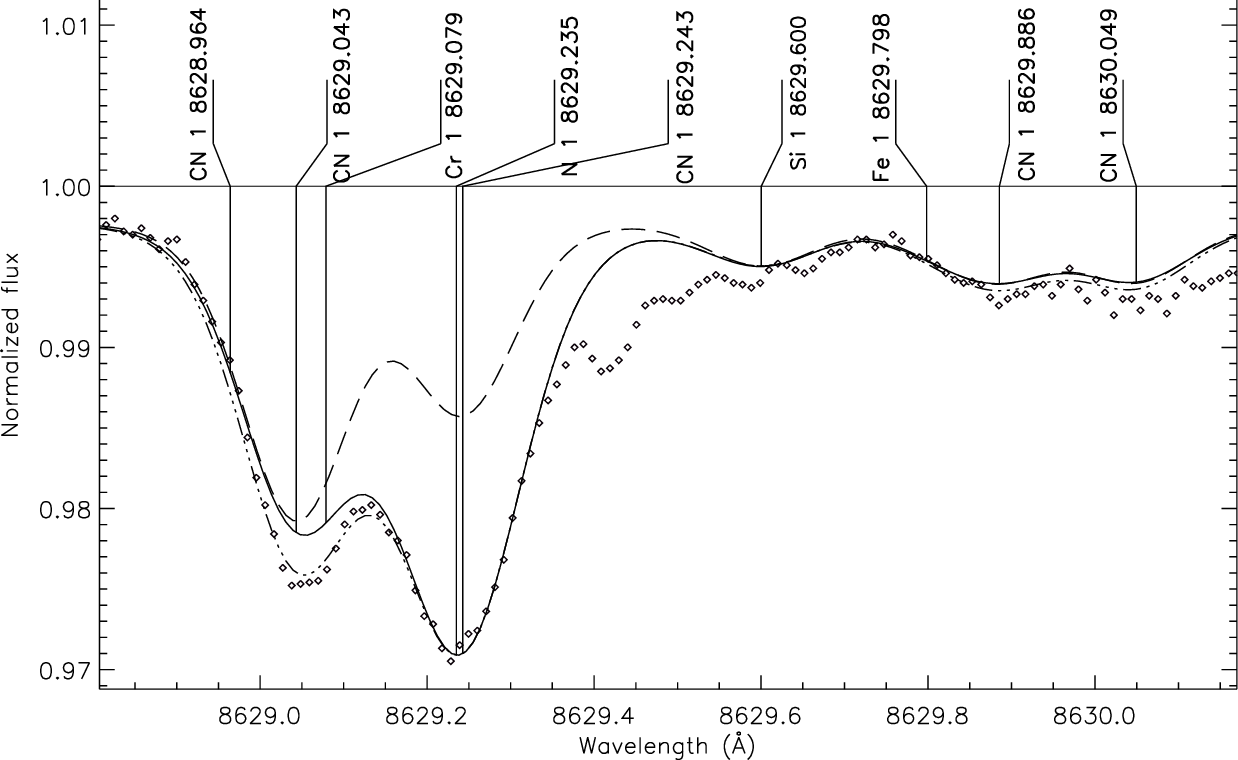}
	\caption{Best NLTE fits (three-dot-dashed curve) to the solar flux spectrum (Kurucz et al. 1984, diamonds) around \ion{N}{1} 7442.28~\AA\ and \ion{N}{1} 8629.23~\AA, obtained with $\eps{N}$, indicated in Table~\ref{tab:sun}, and $\eps{C}$ = 8.53. In each panel: solid curve is the best fit using $\eps{C}$ = 8.43; dashed curve is the synthetic spectrum without the contribution of the \ion{N}{1} line.}
	\label{fig:7442}
\end{figure}

{\bf The \ion{N}{1} 7442.28 and 8629.23~\AA\ lines} are heavily blended, so fitting was carried out over a wide spectrum range, including lines of CN and other elements (Fig.~\ref{fig:7442}). Calculations with fixed carbon abundance $\eps{C}$ = 8.43 showed that, when applying the abundance obtained from the \ion{N}{1} lines: $\eps{N}$ = 7.97 (7442~\AA) and 7.91 (8629~\AA), the theoretical profiles of nearby CN lines appear to be noticeably weaker than the observed ones. The fit improves as the carbon abundance increases by 0.1~dex. In such a case, the nitrogen abundance decreases by 0.06~dex for \ion{N}{1} 7442~\AA\ and by 0.04~dex for the second line. These results are accepted as final and indicated in Table~\ref{tab:sun}.

\begin{figure}  
	\centering
	\includegraphics[width=0.7\columnwidth,clip]{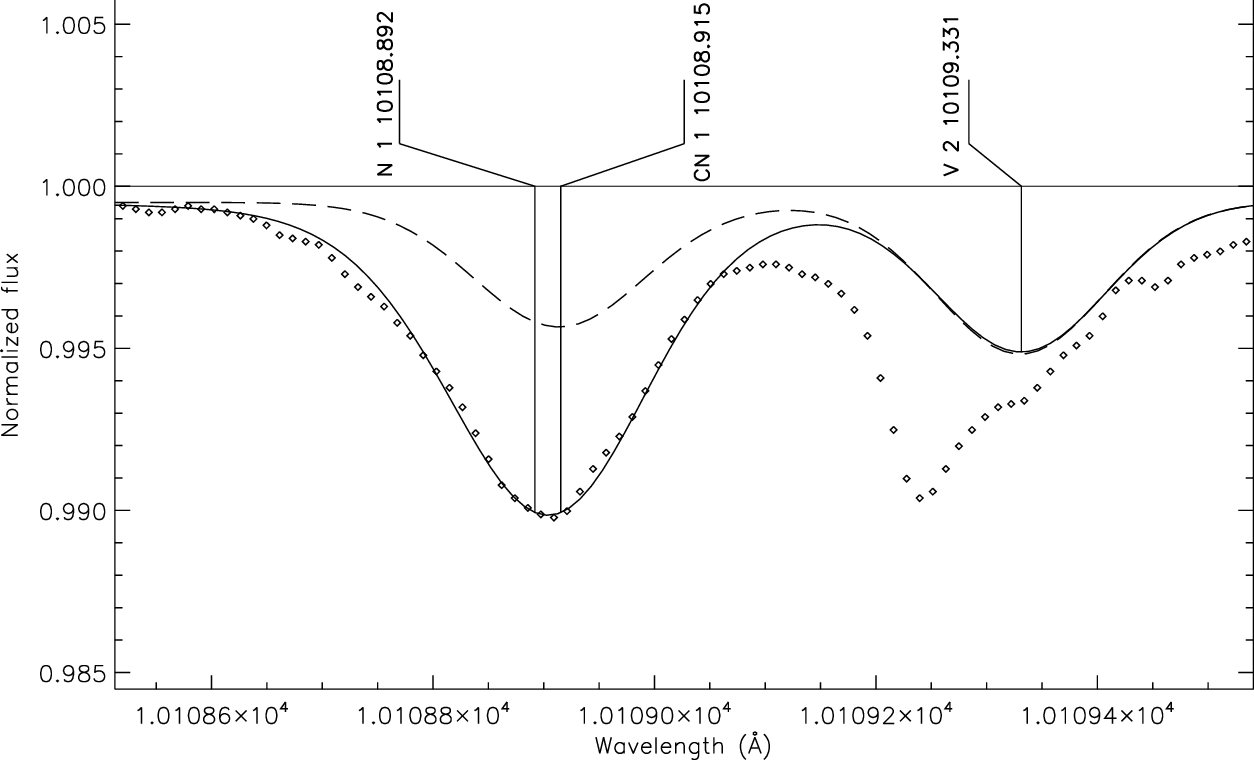}
	\caption{Best NLTE fit (solid curve) to the \ion{N}{1} 10108.89~\AA\ line in the solar flux spectrum (Kurucz et al. 1984, diamonds). The dashed curve is a synthetic spectrum without the contribution of the \ion{N}{1} line.}
	\label{fig:10108}
\end{figure}

{\bf The \ion{N}{1} 10108.89~\AA\ line} overlaps with the CN line. Excluding the CN line from calculations leads to an increase in nitrogen abundance by 0.07~dex. This is the maximum possible uncertainty in the abundance obtained from \ion{N}{1} 10108~\AA.

From five lines of \ion{N}{1}, we obtained the solar average abundance $\eps{N}$(1D NLTE) = 7.92$\pm$0.03. Hereafter, a statistical abundance error is calculated as the dispersion in the single-line measurements around the mean $\sigma = \sqrt{\sum(x-\bar{x})^2/(N_l - 1)}$, where $N_l$ is number of lines. We guess that, in calculations of \citet{Amarsi_n1}, the 3D corrections for individual lines, (3D non-LTE -- 1D non-LTE), are independent of the $EW$-based approach, and we applied them to obtain the NLTE+3D abundances (Table~\ref{tab:sun}) by simply adding our non-LTE abundances and the 3D corrections from \citet[][Table~3]{Amarsi_n1}. The average value is $\eps{N}$(NLTE+3D) = 7.88$\pm$0.03.

Thus, our determinations based on atomic and molecular lines agree within 0.03~dex, and both values agree within error with the solar nitrogen abundance $\eps{\odot,N} = 7.85\pm0.12$, recommended by \citet{lodders21}.

\subsection{Comparisons with literature}\label{sect:sun_comp}

Our abundance from atomic lines agrees within 0.02~dex with $\eps{N}$(3D+NLTE) = 7.86$\pm$0.12  obtained by \cite{Caffau2009}. They performed the 3D LTE calculations for lines of \ion{N}{1}  and then implemented the non-LTE corrections from calculations with a 1D model atmosphere. In the non-LTE calculations, inelastic collision with hydrogen atoms were treated using the \cite{Drawin1969} approximation and applying a scaling factor \kH. In the solar atmosphere, the non-LTE effects for \ion{N}{1} lines are small, and our $\Delta_{\rm NLTE}$ values are consistent within 0.01~dex with the corresponding values of \citet[][Table 4]{Caffau2009}, if \kH\ = 1/3 is adopted. \citet{Caffau2009} used slightly different $gf$-values compared to the data of \cite{n1_nist}. Conversion to new $gf$-values changes the average abundance by less than 0.01~dex.

The difference with \cite{Amarsi_n1} for the abundances, obtained in 1D non-LTE calculations with the same atmospheric model as in this work, has already been noted above. In self-consistent 3D non-LTE calculations, \cite{Amarsi_n1} determined $\eps{N}$(3D~NLTE) = 7.77 from five lines of \ion{N}{1}. For each line, they calculated a less negative non-LTE correction ($\Delta_{\rm NLTE} \simeq -0.01$~dex) compared to ours. Therefore, the lower abundances in their work are mainly due to using equivalent widths and a physically incorrect procedure for subtracting the calculated equivalent widths of the CN lines from the observed equivalent widths of the \ion{N}{1} + CN blends.

Based on the molecular NH and CN lines, the same authors determined $\eps{N}$(3D) = 7.89 \citep{Amarsi_CN}, which is in line with our current results for both atomic and molecular lines.

\cite{Magg2022} use two lines, \ion{N}{1} 8629 and 8683~\AA, with log~$gf$ = 0.006 and 0.162  from own calculations. For the MARCS atmospheric model, they obtained the LTE abundance $\eps{N}$ = 7.88$\pm$0.12. Abundances from individual lines are provided only for the $\ensuremath{\langle\mathrm{3D}\rangle}$ model (their Table~A.1), obtained from the 3D model by spatial and temporal averaging. If we replace oscillator strengths of \cite{Magg2022} with that of \cite{n1_nist}, we obtain the identical abundances from both lines, which are lower by only 0.01~dex than $\eps{N}$ = 7.98$\pm$0.08 indicated by \cite{Magg2022} for the $\ensuremath{\langle\mathrm{3D}\rangle}$ model. Note that, in contrast to \citet{Magg2022}, \citet[][Table~3]{Amarsi_n1} report for the same two \ion{N}{1} lines a noticeably smaller difference between the $\ensuremath{\langle\mathrm{3D}\rangle}$ and MARCS models, 0.014 and 0.018~dex (LTE).

\begin{table*}
	\centering
	\renewcommand{\arraystretch}{1.0}
	\renewcommand{\tabcolsep}{3pt}
	\caption{List of investigated stars, their atmospheric parameters and nitrogen LTE and NLTE abundances.}
	\vspace{3mm}
	\label{tab:stars}
	\begin{tabular}{rcrcrccrccr} %
		\hline\hline \noalign{\smallskip}
		\multicolumn{1}{c}{HD} & Name & \Teff & \lgg & [Fe/H] & $\xi_t$ & Ref & $N_l$ & LTE & \multicolumn{2}{c}{NLTE} \\
		\cline{10-11}
		& & [K]  &      &      &  [\kms] &  & & $\eps{}$ & $\eps{}$ & [N/H] \\
		\noalign{\smallskip}\hline \noalign{\smallskip}
		\ 32115 &              &  7250 & 4.20 &    0.09 & 2.3 & M20 &  8 & ~~7.93(0.07) & ~7.84(0.07) & --0.01  \\
		47105   & $\gamma$ Gem &  9190 & 3.56 &    0.17 & 1.8 & R23 &  8 & ~~8.10(0.06) & ~7.84(0.06) &  --0.01  \\
		\ 48915 & Sirius       &  9850 & 4.30 &    0.52 & 1.8 & M20 & 10 & ~~8.30(0.09) & ~8.07(0.06) &  0.22  \\
		\ 61421 & Procyon      &  6615 & 3.89 & $-$0.01 & 2.0 & R16 &  8 & ~~8.11(0.05) & ~7.97(0.05) &  0.12   \\
		\ 72660 &              &  9700 & 4.10 &    0.67 & 1.8 & M20 &  6 & ~~7.57(0.05) & ~7.41(0.03) & --0.44  \\
		\ 73666 & 40 Cnc       &  9380 & 3.78 &    0.24 & 1.8 & M20 & 12 & ~~8.41(0.12) & ~8.15(0.03) &  0.30  \\
		114330  & $\theta$ Vir &  9600 & 3.61 &    0.31 & 1.4 & R23 & 12 & ~~8.00(0.06) & ~7.77(0.04) &  --0.08   \\
		172167  & Vega         &  9550 & 3.95 & $-$0.41 & 1.8 & M20 &  9 & ~~8.09(0.07) & ~7.82(0.04) & --0.03  \\
		193432  & $\nu$ Cap    & 10200 & 3.88 &    0.12 & 1.0 & R23 &  9 & ~~8.19(0.08) & ~7.93(0.04) &  0.08   \\
		209459  & 21 Peg       & 10400 & 3.55 &    0.05 & 0.5 & M20 & 15 & ~~8.14(0.13) & ~7.89(0.06) &  0.04  \\
		214994  & $o$ Peg      &  9600 & 3.81 &    0.34 & 2.0 & R23 & 15 & ~~8.43(0.15) & ~8.24(0.09) &  0.39   \\
		\noalign{\smallskip}\hline \noalign{\smallskip}
	\end{tabular}

$N_l$ is number of lines. The numbers in parentheses are the statistical errors $\sigma$. Ref: M20 = \citet{mash_a20}, R23 = \citet{Roman2023}, R16 = \citet{Ryabchikova2016}.
\end{table*}

\begin{table}
	\centering
	\renewcommand{\arraystretch}{1.0}
	\renewcommand{\tabcolsep}{7pt}
	\caption{LTE and NLTE abundances obtained from individual lines of \ion{N}{1} in 21~Peg.}
	\vspace{3mm}
	\label{tab:21peg}
	\begin{tabular}{ccrcccc}
		\hline \noalign{\smallskip}
		$\lambda$ & \Eexc & log $gf$ & Transition  & \multicolumn{2}{c}{$\eps{}$} & $\Delta_{\rm NLTE}$   \\
		\cline{5-6}
		[\AA]     &  [eV] &         &          & LTE & NLTE &  [dex] \\
		\noalign{\smallskip} \hline \noalign{\smallskip}
		6482.70 &  11.76 & -0.510 & \eu{3p}{4}{D}{\circ}{7/2} -- \eu{4d}{4}{F}{}{9/2}  & 7.87 &  7.79 & -0.08 \\
		6644.96 &  11.76 & -0.858 & \eu{3p}{4}{D}{\circ}{7/2} -- \eu{5s}{4}{P}{}{5/2}  & 7.99 &  7.90 & -0.09 \\
		6722.61 &  11.84 & -0.714 & \eu{3p}{4}{P}{\circ}{5/2} -- \eu{4d}{4}{D}{}{7/2}  & 7.95 &  7.86 & -0.09 \\
		7442.30 &  10.33 & -0.401 & \eu{3s}{4}{P}{}{3/2} -- \eu{3p}{4}{S}{\circ}{3/2}  & 8.13 &  7.86 & -0.27 \\
		7468.31 &  10.34 & -0.183 & \eu{3s}{4}{P}{}{5/2} -- \eu{3p}{4}{S}{\circ}{3/2}  & 8.11 &  7.81 & -0.30 \\
		8184.86 &  10.33 & -0.305 & \eu{3s}{4}{P}{}{3/2} -- \eu{3p}{4}{P}{\circ}{5/2}  & 8.23 &  7.91 & -0.32 \\
		8188.01 &  10.33 & -0.298 & \eu{3s}{4}{P}{}{1/2} -- \eu{3p}{4}{P}{\circ}{3/2}  & 8.23 &  7.90 & -0.33 \\
		8567.74 &  10.67 & -0.670 & \eu{3s}{2}{P}{}{1/2} -- \eu{3p}{2}{P}{\circ}{3/2}  & 8.16 &  8.05 & -0.11 \\
		8629.24 &  10.69 &  0.077 & \eu{3s}{2}{P}{}{3/2} -- \eu{3p}{2}{P}{\circ}{3/2}  & 8.05 &  7.90 & -0.15 \\
		8683.40 &  10.33 &  0.105 & \eu{3s}{4}{P}{}{3/2} -- \eu{3p}{4}{D}{\circ}{5/2}  & 8.38 &  7.90 & -0.48 \\
		8686.15 &  10.33 & -0.284 & \eu{3s}{4}{P}{}{1/2} -- \eu{3p}{4}{D}{\circ}{3/2}  & 8.21 &  7.90 & -0.31 \\
		8703.25 &  10.33 & -0.310 & \eu{3s}{4}{P}{}{1/2} -- \eu{3p}{4}{D}{\circ}{1/2}  & 8.14 &  7.87 & -0.27 \\
		8711.70 &  10.33 & -0.233 & \eu{3s}{4}{P}{}{3/2} -- \eu{3p}{4}{D}{\circ}{3/2}  & 8.17 &  7.89 & -0.28 \\
		8718.84 &  10.34 & -0.349 & \eu{3s}{4}{P}{}{5/2} -- \eu{3p}{4}{D}{\circ}{5/2}  & 8.11 &  7.85 & -0.26 \\
		8728.90 &  10.33 & -1.067 & \eu{3s}{4}{P}{}{3/2} -- \eu{3p}{4}{D}{\circ}{1/2}  & 8.05 &  7.85 & -0.20 \\
		\noalign{\smallskip}\hline \noalign{\smallskip}
	\end{tabular}
\end{table}

\section{Abundance of nitrogen in the A-stars and Procyon}\label{sect:sample}
\subsection{Stellar sample, observations and atmospheric parameters}

In this study, we analyse lines of \ion{N}{1} in the stars with \Teff\ $<$ 11\,000~K, which were selected from our previous works \citep{mash_a20,Roman2023,Ryabchikova2016}. We use the same atmospheric parameters determined, for each star, by several methods and the same observed spectra obtained in a wide spectral range with R $\ge$ 60\,000 on the CFHT (Canada-France-Hawaii Telescope) and VLT2 (Very Large Telescope) telescopes. More details about the observed spectra and the determination of atmospheric parameters can be found in the cited papers.
The sample stars and their atmospheric parameters, \Teff, \lgg, iron abundance [Fe/H]\footnote{for any two elements X and Y: [X/Y] = log$(N_{\rm X}/N_{\rm Y})_{star} - \log (N_{\rm X}/N_{\rm Y})_{Sun}$.}, and $\xi_t$ are listed in Table~\ref{tab:stars}.

According to our previous determinations \citep{mash_a20,Roman2023}, in five stars, the non-LTE abundances of He, C, O, Na, Mg, Si, Ca, Ti, and Fe are consistent with the solar ones within 0.1~dex. We call them as superficially normal stars. The star HD~61421 (Procyon) is a reference F-type star with a solar chemical composition,as confirmed by \citet{Adibekyan2020}. HD~48915 (Sirius), HD~72660, HD~114330 ($\theta$ Vir) and HD~214994 ($o$ Peg) reveal enhanced metal lines, and they are referred to as metllic-lined (Am) stars. HD~172167 (Vega) belongs to $\lambda$~Boo-type stars, which have a deficiency of Mg, Al, Si, S, Mn, Fe, Ni in their atmospheres, but close-to-solar abundances of volatile elements C, N, O.

\subsection{Nitrogen abundance determinations}	

\begin{figure*}  
	\centering
	\includegraphics[width=0.7\columnwidth,clip]{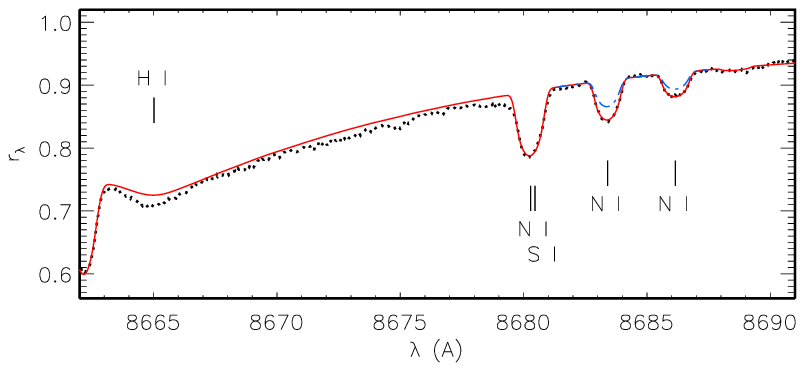}
	\caption{\ion{N}{1} 8683 and 8686~\AA\ lines in Vega spectrum (dotted curve) and the best fit (solid curve) obtained with $\eps{N}$(NLTE) = 7.83 and 7.76, respectively. For comparison, the dashed-dotted curve shows the theoretical spectrum from LTE calculations with the same nitrogen abundances.}
	\label{fig:vega}
\end{figure*}

\begin{figure}  
	\centering
	\includegraphics[width=0.55\columnwidth,clip]{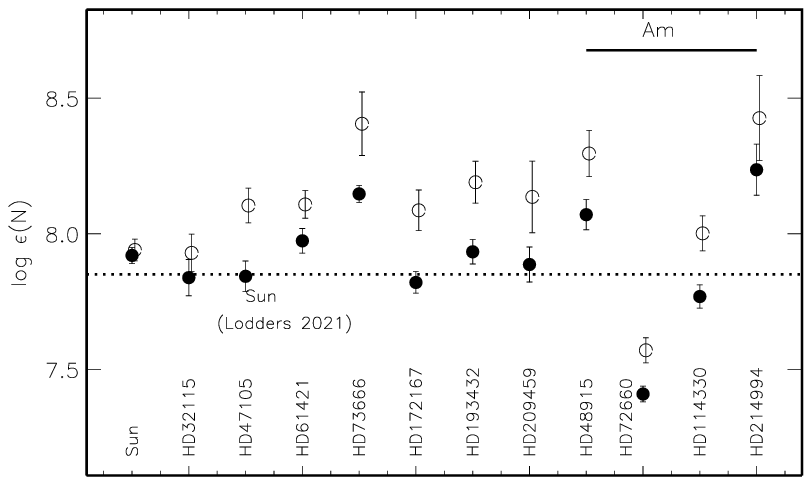}
	\caption{Nitrogen NLTE (filled circles) and LTE (open circles) abundances $\eps{N}$ of the sample stars. The dotted line corresponds to the solar abundance recommended by Lodders (2021).}
	\label{fig:stars}
\end{figure}

Due to the higher temperatures of A-type stars, more lines of \ion{N}{1} can be measured in their spectra compared to the solar one. Lines used in abundance determinations and their atomic parameters are listed in Table~\ref{tab:21peg}. Eight of them, with $\lambda_0 \ge$ 8567~\AA, are located in the wings of the hydrogen Paschen lines. For six stars from  \citet{mash_a20} and for Procyon, the LTE and non-LTE abundances from individual lines were determined using the synthetic spectrum method and the codes \textsc{synthV\_NLTE} and BinMag, similarly to the analysis of the solar spectrum. Quality of fitting the \ion{N}{1} 8683 and 8686~\AA\ lines in the observed spectrum of Vega is shown in Fig.~\ref{fig:vega}. For the remaining four stars, the LTE abundances for individual lines were taken from Romanovskaya et al. (2023), and the non-LTE abundances were obtained by adding the non-LTE corrections calculated in this study for given atmospheric parameters.
The LTE and non-LTE abundances for individual lines are presented in Table~\ref{tab:21peg} for 21~Peg and Appendix (Table~A.1) for the remaining stars. The average nitrogen abundances for all stars are presented in Table~\ref{tab:stars} and in Fig.~\ref{fig:stars}. The [N/H] values were calculated using the solar abundance $\eps{\odot,N}$ = 7.85, recommended by Lodders (2021).

For each star, non-LTE leads to a decreased statistical abundance error. For several stars, such a decrease is significant. For example, for HD~73666, $\sigma$ = 0.12~dex in the LTE case, but $\sigma$ = 0.03~dex in the non-LTE calculations.

Our results show that superficially normal A stars have solar nitrogen abundances. Non-LTE removes the enhancements relative to the solar N abundance obtained in the LTE analysis. The excess [N/H](NLTE) = 0.30 in HD~73666 can be understood because this star is a member of the Praesepe open cluster, which is characterised by a supersolar metallicity of [Fe/H] = 0.16 (Netopil et al. 2022). Mashonkina et al. (2020) determined [Fe/H] = 0.24 for HD~73666. Thus, each of the normal A stars retains in its atmosphere the nitrogen abundance given to it at birth.

For Vega, the nitrogen abundance is found to be close-to-solar, as previously determined for C and O (Mashonkina et al., 2020). This is consistent with the present-day ideas about the nature of $\lambda$~Boo-type stars, which accrete circumstellar gas with low abundances of the elements Mg, Al, Si, S, Mn, Fe, Ni, involved in the formation of dust (Venn, Lambert, 1990), but have close-to-solar abundances of volatile elements C, N, O.

Four Am stars show a wide variation in nitrogen abundance, from a deficiency of [N/H] = $-0.44$ to an excess of [N/H] = 0.39. According to our determinations for a large number of elements (Mashonkina et al., 2020; Romanovskaya et al., 2023; Mashonkina, 2024), a similar and even greater scatter is observed in these stars only for Ca ([Ca/H] from $-0.22$ to 0.41) and Sc ([Sc/H] from $-0.80$ to 0.49).

\subsection{Uncertainties in non-LTE abundances}

\begin{table}
	\centering
	\caption{Error estimates for NLTE calculations of the \ion{N}{1} lines in 21~Peg.}
	\label{tab:uncertainties}
	\begin{tabular}{llrr} %
		\hline\hline \noalign{\smallskip}
		& \multicolumn{3}{r}{Changes in $\eps{NLTE}$ (dex)} \\
		\cline{3-4}
		& & 7442\,\AA & 8683\,\AA \\
		\noalign{\smallskip}\hline \noalign{\smallskip}
		LTE & $-\Delta_{\rm NLTE}$ & 0.27 & 0.48  \\
		Photoionization & $\sigma_{\rm RBF}$ & $-0.02$ &  $-0.03$ \\
		\multicolumn{4}{l}{(hydrogenic cross-sections)} \\
		Electron-impact excitation & $\sigma_{\rm CBB}$ & +0.05 & +0.09 \\
		\multicolumn{4}{l}{(approximate formulas for all transitions)} \\
		\Teff\ (--200~K) & $\sigma$(\Teff) & $-0.03$ & $-0.02$ \\
		\lgg\ (+0.1)    & $\sigma$(\lgg) & $0.01$ & $0.02$   \\
		$\xi_t$ (+0.4~\kms)  & $\sigma (\xi_t)$ & $0.00$ & $-0.01$  \\
		\noalign{\smallskip}\hline \noalign{\smallskip}
		\multicolumn{4}{l}{{\bf Note:} 0.00 means smaller than 0.01~dex, in absolute value. }
	\end{tabular}
\end{table}

Observational errors due to the uncertainty in the continuum level and the choice of spectral range for fitting the \ion{N}{1} lines, as well the $gf$ errors are included in the statistical error $\sigma$. We have the opportunity to estimate the observational error caused by using different spectra for one star.
    For Vega, we used the spectrum obtained by A.~Korn with the Fibre Optics Cassegrain Echelle Spectrograph (FOCES) on the 2.2-m telescope at the Calar Alto Observatory (Spain); R $\simeq$ 40\,000. Another spectrum was taken by Takeda et al. (2007) with the HIgh-Dispersion Echelle Spectrograph (HIDES) on the 1.88-m telescope at Okayama Observatory\footnote{\tt http://pasj.asj.or.jp/v59/n1/590122/590122-frame.html} (Japan); R $\simeq$ 100\,000. In the second spectrum, in the range 7620-8809~\AA, it was possible to measure five lines of \ion{N}{1}. They are all located in the wings of the hydrogen lines of the Paschen series. The difference in abundance between using FOCES and HIDES spectra is $-0.02\pm0.05$. Thus, even for such heavily blended lines, the synthetic spectrum method allows us to obtain a reliable nitrogen abundance.

Using the star 21~Peg and two lines, \ion{N}{1} 7442 and 8683~\AA, we estimated the uncertainties in the obtained non-LTE abundances due to uncertainties in the non-LTE method and atmospheric parameters. Test calculations were carried out with hydrogenic photoionization cross-sections instead of quantum-mechanical ones from the Opacity Project and with the approximate formula of van Regemorter (1962) and $\Omega$ = 1 to calculate the electron-impact excitation rates in all b-b transitions, and not just those that are missing in the calculations of Wang et al. (2014). For 21~Peg, Fossati et al. (2009) estimated the uncertainties in \Teff, \lgg\ and $\xi_t$ as 200~K, 0.1~dex and 0.4~\kms. Table~\ref{tab:uncertainties} summarizes the results of our tests.

The non-LTE results are most sensitive to changes in collisional rates, since we are dealing with spectral lines formed between highly excited levels, with an energy separation comparable to the average kinetic energy of electrons. The use of approximate formulas leads to weakened non-LTE effects, but the non-LTE corrections for the \ion{N}{1} lines still remain large in absolute value.
    
\subsection{Comparisons with literature}

\begin{figure}  
	\centering
	\includegraphics[width=0.45\columnwidth,clip]{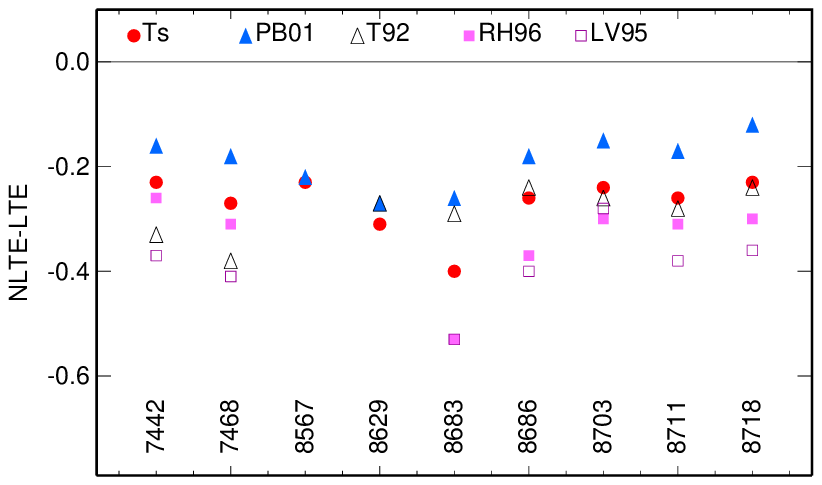}
	\caption{Non-LTE abundance corrections for individual lines of \ion{N}{1}, calculated by different authors with a model of Vega's atmosphere. Line wavelengths in angstroms are indicated at the bottom of the panel. Various symbols correspond to this work (Ts, filled circles), Przybilla and Butler (2001, PB01, filled triangles), Takeda (1992, T92, empty triangles), \citet[][RH96, filled squares]{n1_sun1996} and \citet[][LV95, empty squares]{Lemke95}.}
	\label{fig:dnlte}
\end{figure}

Of all the sample stars, Vega has largest set of nitrogen non-LTE abundance determinations in the literature. In all the works, very similar \Teff\ from 9500 to 9550~K and \lgg\ from 3.9 to 4.0 were employed, so the differences in results cannot be explained by differences in atmospheric parameters. Using the equivalent widths, even for the \ion{N}{1} lines in the hydrogen line wings, \cite{takeda1992} and \cite{n1_sun1996} obtained $\eps{N} \sim$ 7.5 (solar metallicity model) and $\eps{N}$ = 7.76 (7.67 after applying the NIST $gf$-values). \cite{Lemke95} and \cite{Przybilla_n1} used the synthetic spectrum method and determined $\eps{N} \sim$ 7.6 and $\eps{N}$ = 7.67. The use of the NIST $gf$-values would change the average abundances in two latter papers by less than 0.01~dex. The non-LTE corrections for individual lines reported in different papers are compared in Fig.~\ref{fig:dnlte}.

\cite{takeda1992} and \cite{Lemke95} note the sensitivity of their results to variations in the background opacity around Lyman $\alpha$. 
In our {\sc detail} code, opacity is calculated taking into account the quasi-molecular Lyman $\alpha$ satellites following \cite{Allard1998}.
 This may explain the weaker non-LTE effects for \ion{N}{1} and the higher obtained nitrogen abundances in our work compared to \cite{takeda1992}, \cite{Lemke95}, \cite{n1_sun1996}.

We see substantial discrepancies with the results of Przybilla and Butler (2001). They are not only due to differences in the non-LTE methods. Przybilla and Butler (2001) calculated systematically less negative non-LTE corrections than ours, by 0.1~dex, on average, but obtained a lower nitrogen abundance, by 0.15~dex. We deduce from this that the LTE abundance in their work is substantially lower than ours. This is surprising since we use the same FOCES spectrum, the same atmospheric parameters and close together $gf$-values.

For seven stars in common with our sample, the non-LTE abundances were determined by \cite{takeda2018} using the \ion{N}{1} 7468~\AA\ line. For Procyon, our results are in line with that of \cite{takeda2018}. For five A-type stars, with \Teff\ and \lgg, which agree with our values within 150~K and 0.1~dex, \cite{takeda2018} determined systematically lower NLTE abundances compared to ours, by 0.08--0.12~dex. This can be due to applying the non-LTE method of \cite{takeda1992} that overestimates the non-LTE effects for \ion{N}{1}, as seen in Fig.~\ref{fig:dnlte}. For $o$~Peg, the N abundance in \citet{takeda2018} is lower than ours, by 0.26~dex, probably, due to the lower \lgg\ in their work.

We have eight stars in common with Romanovskaya et al. (2023), who used the non-LTE method developed by Lyubimkov et al. (2011). Compared to our results, Romanovskaya et al. (2023) calculated stronger non-LTE effects on the \ion{N}{1} lines in the stars with \Teff\ $>$ 9000~K. The difference between the two papers depends mainly on stellar \Teff. For example, with the model 9190/3.6, \cite{Roman2023} calculated more negative non-LTE abundance corrections than ours, by 0.05~dex and 0.16~dex for \ion{N}{1} 7442 and 8683~\AA, respectively. While in the 10200/3.9 model, the corresponding values are 0.20~dex and 0.28~dex. Romanovskaya et al. (2023) obtained the lower average non-LTE abundances compared to our results, by 0.09--0.36~dex for different stars, and a nitrogen deficiency of [N/H] = $-0.11$ to $-0.32$ for superficially normal A stars, in which the abundances of the elements from He to Fe are close to the solar ones. It is difficult to imagine a mechanism that leads to anomalous nitrogen abundance in a star's atmosphere without affecting the abundances of other elements. An argument in favor of our results is that for these same stars the nitrogen non-LTE abundances appear to be solar.

\section{Conclusions}\label{conclusions}

A new model of the nitrogen atom has been constructed, which includes energy levels of \ion{N}{1} from laboratory measurements, as presented in NIST, and also energy levels predicted by R. Kurucz in the atomic structure calculations for \ion{N}{1}. Radiative and collisional rates for b-b and b-f transitions are calculated using the most up-to-date available so far atomic data on photoionization cross-sections, electron-impact excitation cross-sections, and rate coefficients for inelastic processes in collisions with hydrogen atoms.

Non-LTE calculations for \ion{N}{1} were carried out with the classical MARCS solar model atmosphere. Non-LTE leads to strengthened lines of \ion{N}{1} and negative non-LTE abundance corrections, but the effects are small, that is, $\Delta_{\rm NLTE}$ does not exceed 0.05~dex, in absolute value. Using the synthetic spectrum method, we derived the solar non-LTE abundance $\eps{N}$(1D NLTE) = 7.92$\pm$0.03 from five atomic lines. Applying the 3D corrections for individual lines from calculations of Amarsi et al. (2020), we obtained $\eps{N}$(NLTE+3D) = 7.88$\pm$0.03. This value is consistent with $\eps{N}$(3D) = 7.89, obtained by Amarsi et al. (2021) from the NH and CN molecular lines, and with $\eps{N}$ = 7.91, obtained by Ryabchikova et al. (2022) from the CN lines and corrected for 3D effects in accordance with Amarsi et al. (2021). Our results are consistent within the error bars with the solar abundance $\eps{\odot,N} = 7.85\pm0.12$, recommended by Lodders (2021).

Non-LTE calculations were performed for 11 unevolved F-A type stars with reliably determined atmospheric parameters. It is shown that negative non-LTE abundance corrections increase in absolute value with increasing \Teff, so that $\Delta_{\rm NLTE}$ reaches $-0.48$~dex for \ion{N}{1} 8683~\AA\ in the model 10400/3.55. The nitrogen non-LTE abundances were determined based on the observed high-quality spectra. For superficially normal A stars, non-LTE removes the excess relative to the solar nitrogen abundance obtained in LTE analysis. As for elements from He to Fe, the nitrogen non-LTE abundance of these stars differs from the solar abundance by no more than 0.1~dex. A solar abundance was also obtained for nitrogen in Vega, a star of the $\lambda$~Boo type. Thus, we have expanded the list of elements in a detailed analysis of the chemical composition of A-type stars -- a project whose results were presented earlier by Mashonkina et al. (2020), Sitnova et al. (2022), Romanovskaya et al. (2023), Mashonkina (2024). Four Am stars show a variation in nitrogen abundance, from a deficiency of [N/H] = $-0.44$ to an excess of [N/H] = 0.39.

If we do not take into account HD~73666, which was born in the open cluster with supersolar metallicity, but add Vega to the rest of the superficially normal A stars, for which no deviation in the nitrogen abundance is expected from its abundance in the matter from which the star was formed, then the average for five stars amounts to $\eps{N}$ = 7.86$\pm$0.04. Nieva and Przybilla (2012) proposed the cosmic nitrogen abundance standard $\eps{N} = 7.79\pm0.04$, as determined from lines of \ion{N}{2} in a sample of early B-type stars. For comparison, they cited the abundance obtained from  analysis of the interstellar UV lines of \ion{N}{1}, $\eps{N} = 7.79\pm0.03$ \citep{Meyer1997}. The difference in N abundance between the Sun, A--B stars and the interstellar gas is 0.09~dex or $\sim2\sigma$, and this is an evidence for  the reliability of modern methods for studying the chemical composition of the galactic matter.

\acknowledgments
LM acknowledges the Russian Science Foundation (grant 23-12-00134) for a partial support of this study (Sects.~\ref{sect:method} and \ref{sect:sun_comp}). This study made use of the NIST, VALD and ADS\footnote{http://adsabs.harvard.edu/abstract\_service.html} databases.

\clearpage
\begin{table}
	\centering
	\renewcommand{\arraystretch}{1.0}
	\renewcommand{\tabcolsep}{4pt}
	\caption{Appendix A1: LTE and NLTE abundances from individual lines of \ion{N}{1} in the sample stars.}
	\vspace{3mm}
	\begin{tabular}{crccc|crccc}
		\hline \noalign{\smallskip}
		$\lambda$ & log $gf$ & \multicolumn{2}{c}{$\eps{}$} & $\Delta_{\rm NLTE}$ & $\lambda$ & lg $gf$ & \multicolumn{2}{c}{$\eps{}$} & $\Delta_{\rm NLTE}$  \\
		\cline{3-4} 
		\cline{8-9} 
		[\AA]     &         &   LTE & NLTE &  [dex] & [\AA]     &         &   LTE & NLTE &  [dex] \\
		\noalign{\smallskip} \hline \noalign{\smallskip}		
\multicolumn{5}{l}{Procyon     } &           \multicolumn{5}{|l}{Sirius}     \\
7442.28 & -0.401 &  8.09 &  7.99 & -0.11 &   7442.30 & -0.401 &  8.21 &  8.03 & -0.18 \\
8184.86 & -0.305 &  8.13 &  8.01 & -0.12 &   7468.31 & -0.183 &  8.17 &  7.96 & -0.21 \\
8188.01 & -0.298 &  8.14 &  8.01 & -0.12 &   8184.86 & -0.305 &  8.25 &  8.03 & -0.22 \\
8216.34 &  0.138 &  8.14 &  7.97 & -0.17 &   8567.74 & -0.670 &  8.27 &  8.16 & -0.11 \\
8683.40 &  0.105 &  8.15 &  7.98 & -0.17 &   8629.24 &  0.077 &  8.32 &  8.13 & -0.19 \\
8703.25 & -0.310 &  8.03 &  7.91 & -0.12 &   8683.40 &  0.105 &  8.46 &  8.06 & -0.40 \\
8711.70 & -0.233 &  8.02 &  7.89 & -0.13 &   8686.15 & -0.284 &  8.32 &  8.08 & -0.24 \\
8718.84 & -0.349 &  8.15 &  8.02 & -0.13 &   8703.25 & -0.310 &  8.31 &  8.09 & -0.22 \\
\multicolumn{5}{l|}{$\gamma$ Gem}        &   8711.70 & -0.233 &  8.32 &  8.08 & -0.24 \\
7442.30 & -0.401 &  8.03 &  7.80 & -0.23 &   8718.84 & -0.349 &  8.26 &  8.05 & -0.21 \\
7468.31 & -0.183 &  8.03 &  7.76 & -0.27 & \multicolumn{5}{|l}{$\theta$ Vir}   \\
8184.86 & -0.305 &  8.20 &  7.94 & -0.26 &   7442.30 & -0.401 &  7.92 &  7.70 & -0.22 \\
8629.24 &  0.077 &  8.06 &  7.82 & -0.24 &   7468.31 & -0.183 &  7.95 &  7.70 & -0.25 \\
8683.40 &  0.105 &  8.17 &  7.81 & -0.36 &   8184.86 & -0.305 &  8.00 &  7.75 & -0.25 \\
8686.15 & -0.284 &  8.12 &  7.86 & -0.26 &   8188.01 & -0.298 &  8.05 &  7.80 & -0.25 \\
8703.25 & -0.310 &  8.09 &  7.87 & -0.22 &   8567.74 & -0.670 &  7.98 &  7.81 & -0.17 \\
8711.70 & -0.233 &  8.10 &  7.86 & -0.24 &   8629.24 &  0.077 &  7.97 &  7.75 & -0.22 \\
\multicolumn{5}{l|}{HD~32115}            &   8683.40 &  0.105 &  8.14 &  7.82 & -0.32 \\
6644.96 & -0.858 &  7.94 &  7.83 & -0.11 &   8686.15 & -0.284 &  8.06 &  7.82 & -0.24 \\
7442.30 & -0.401 &  7.93 &  7.85 & -0.08 &   8703.25 & -0.310 &  7.97 &  7.76 & -0.21 \\
7468.31 & -0.183 &  7.85 &  7.84 & -0.01 &   8711.70 & -0.233 &  7.97 &  7.73 & -0.24 \\
8629.24 &  0.077 &  7.92 &  7.83 & -0.09 &   8718.84 & -0.349 &  7.98 &  7.76 & -0.22 \\
8683.40 &  0.105 &  7.87 &  7.73 & -0.14 &   8728.90 & -1.067 &  7.98 &  7.80 & -0.18 \\
8703.25 & -0.310 &  7.99 &  7.89 & -0.10 & \multicolumn{5}{|l}{$o$ Peg}   \\
8711.70 & -0.233 &  7.86 &  7.76 & -0.10 &   6482.70 & -0.510 &  8.06 &  8.00 & -0.06 \\
8718.84 & -0.349 &  8.04 &  7.94 & -0.10 &   6644.96 & -0.858 &  8.31 &  8.25 & -0.06 \\
\multicolumn{5}{l|}{HD 73666}            &   6722.61 & -0.714 &  8.16 &  8.10 & -0.06 \\
6482.70 & -0.510 &  8.23 &  8.19 & -0.04 &   7442.30 & -0.401 &  8.40 &  8.18 & -0.22 \\
7442.30 & -0.401 &  8.34 &  8.11 & -0.23 &   7468.31 & -0.183 &  8.41 &  8.17 & -0.24 \\
7468.31 & -0.183 &  8.35 &  8.08 & -0.27 &   8184.86 & -0.305 &  8.51 &  8.29 & -0.22 \\
8184.86 & -0.305 &  8.40 &  8.13 & -0.27 &   8188.01 & -0.298 &  8.55 &  8.33 & -0.22 \\
8567.74 & -0.670 &  8.37 &  8.19 & -0.18 &   8567.74 & -0.670 &  8.40 &  8.27 & -0.13 \\
8629.24 &  0.077 &  8.43 &  8.13 & -0.30 &   8629.24 &  0.077 &  8.39 &  8.17 & -0.22 \\
8683.40 &  0.105 &  8.65 &  8.16 & -0.49 &   8683.40 &  0.105 &  8.70 &  8.39 & -0.31 \\
8686.15 & -0.284 &  8.44 &  8.15 & -0.29 &   8686.15 & -0.284 &  8.51 &  8.30 & -0.21 \\
8703.25 & -0.310 &  8.41 &  8.16 & -0.25 &   8703.25 & -0.310 &  8.44 &  8.25 & -0.19 \\
8711.70 & -0.233 &  8.40 &  8.14 & -0.26 &   8711.70 & -0.233 &  8.49 &  8.30 & -0.19 \\
8718.84 & -0.349 &  8.38 &  8.14 & -0.24 &   8718.84 & -0.349 &  8.39 &  8.21 & -0.18 \\
8728.90 & -1.067 &  8.33 &  8.17 & -0.16 &   8728.90 & -1.067 &  8.30 &  8.18 & -0.12 \\
\noalign{\smallskip}\hline \noalign{\smallskip}
\end{tabular}
\end{table}

\begin{table}
\centering
\renewcommand{\arraystretch}{1.0}
\renewcommand{\tabcolsep}{4pt}
\caption{Appendix A1, continued.}
\vspace{3mm}
\begin{tabular}{crccc|crccc}
\hline \noalign{\smallskip}
$\lambda$ & lg $gf$ & \multicolumn{2}{c}{$\eps{}$} & $\Delta_{\rm NLTE}$ & $\lambda$ & lg $gf$ & \multicolumn{2}{c}{$\eps{}$} & $\Delta_{\rm NLTE}$  \\
\cline{3-4}
\cline{8-9}
[\AA]     &         &   LTE & NLTE &  [dex] & [\AA]     &         &   LTE & NLTE &  [dex] \\
\noalign{\smallskip} \hline \noalign{\smallskip}
\multicolumn{5}{l}{$\nu$ Cap}            & \multicolumn{5}{|l}{Vega}      \\
7442.30 & -0.401 &  8.09 &  7.84 & -0.25 &   7442.30 & -0.401 &  8.01 &  7.79 & -0.22 \\
7468.31 & -0.183 &  8.16 &  7.87 & -0.29 &   7468.31 & -0.183 &  8.06 &  7.80 & -0.26 \\
8567.74 & -0.670 &  8.13 &  7.97 & -0.16 &   8567.74 & -0.670 &  8.13 &  7.90 & -0.23 \\
8629.24 &  0.077 &  8.17 &  7.96 & -0.21 &   8629.24 &  0.077 &  8.14 &  7.83 & -0.31 \\
8683.40 &  0.105 &  8.34 &  7.97 & -0.37 &   8683.40 &  0.105 &  8.22 &  7.83 & -0.39 \\
8686.15 & -0.284 &  8.21 &  7.94 & -0.27 &   8686.15 & -0.284 &  8.01 &  7.76 & -0.25 \\
8703.25 & -0.310 &  8.18 &  7.94 & -0.24 &   8703.25 & -0.310 &  8.06 &  7.83 & -0.23 \\
8711.70 & -0.233 &  8.22 &  7.97 & -0.25 &   8711.70 & -0.233 &  8.07 &  7.82 & -0.25 \\
8718.84 & -0.349 &  8.16 &  7.92 & -0.24 &   8718.84 & -0.349 &  8.03 &  7.81 & -0.22 \\
\multicolumn{5}{l|}{HD 72660}            &  \multicolumn{5}{l}{ }  \\
8629.24 &  0.077 &  7.50 &  7.36 & -0.14 &  \multicolumn{5}{|l}{ }  \\
8683.40 &  0.105 &  7.62 &  7.41 & -0.21 &  \multicolumn{5}{|l}{ }  \\
8686.15 & -0.284 &  7.62 &  7.45 & -0.17 &  \multicolumn{5}{|l}{ }  \\
8703.25 & -0.310 &  7.55 &  7.41 & -0.14 &  \multicolumn{5}{|l}{ }  \\
8711.70 & -0.233 &  7.55 &  7.40 & -0.15 &  \multicolumn{5}{|l}{ }  \\
8718.84 & -0.349 &  7.57 &  7.42 & -0.15 &  \multicolumn{5}{|l}{ }  \\
\noalign{\smallskip}\hline \noalign{\smallskip}
\end{tabular}
\end{table}


\clearpage
\begin{thebibliography}{}
\bibitem[Adibekyan et al.(2020)]{Adibekyan2020}
V. Adibekyan, S. G. Sousa, N. C. Santos, P. Figueira, C. Allende Prieto, E. Delgado Mena, J. I. Gonz{\'a}lez Hern{\'a}ndez, P. de Laverny, A. Recio-Blanco, T. L. Campante, M. Tsantaki, A. A. Hakobyan, M. Oshagh, J. P. Faria, M. Bergemann, G. Israelian, T. Boulet, \aap\ {\bf 642}, A182 (2020).
\bibitem[Alexeeva and Mashonkina(2015)]{c_sun2015}
S. A. Alexeeva and L. I. Mashonkina, \mnras\ {\bf 453}, 1619 (2015)
\bibitem[Allard et al. (1998)]{Allard1998}
N. F. Allard, I. Drira, M. Gerbaldi, J. Kielkopf, A. Spielfiedel, \aap\ {\bf 335}, 1124 (1998)
\bibitem[Amarsi and Barklem(2019)]{Amarsi_n1_hyd}
A. M. Amarsi and  P. S. Barklem, \aap\ {\bf  625}, A78 (2019).
\bibitem[Amarsi et al.(2020)]{Amarsi_n1}
A. M. Amarsi, N. Grevesse, J. Grumer, M. Asplund, P. S. Barklem, R. Collet, \aap\ {\bf 636}, A120 (2020).
\bibitem[Amarsi et al.(2021)]{Amarsi_CN}
A. M. Amarsi, N. Grevesse, M. Asplund, R. Collet, \aap\ {\bf 656}, A113 (2021).
\bibitem[Anders and Grevesse(1989)]{cosmos89}
E. Anders and N. Grevesse, Geochimica et Cosmochimica Acta {\bf 53} , 197 (1989).
\bibitem[Asplund et al.(2021)]{Asplund2021}
M. Asplund, A. M. Amarsi, N. Grevesse, \aap\ {\bf 653}, A141 (2021).
\bibitem[Bagnulo et al., 2003]{Bagnulo2003}
S. Bagnulo, E. Jehin, C. Ledoux, R. Cabanac, C. Melo, R. Gilmozzi, ESO Paranal Science Operations Team, ESO Messenger {\bf 114}, 10 (2003).
\bibitem[Bi{\'e}mont et al.(1990)]{sun1990}
E. Bi{\'e}mont, C. Froese Fischer, M. Godefroid, N. Vaeck, A. Hibbert, 3rd International Colloquium of the Royal Netherlands Academy of Arts and Sciences, 59 (1990)
\bibitem[Brooke et al.(2014)]{CN}
J. S. A. Brooke, R. S. Ram, C. M. Western, G. Li, D. W. Schwenke,  P. F. Bernath, Astrophys. J. Suppl. Ser. {\bf 210}, 23 (2014).
\bibitem[Butler, 1984]{butler84}
K. Butler, Ph.D. Thesis, University of London (1984).
\bibitem[Caffau et al.(2009)]{Caffau2009}
E. Caffau, E. Maiorca, P. Bonifacio, R. Faraggiana, M. Steffen, H.-G. Ludwig, I. Kamp, M. Busso, \aap\ {\bf 498}, 877 (2009).
\bibitem[Cunto et al., 1993]{topbase}
W. Cunto, C. Mendoza, F. Ochsenbein, C. J. Zeippen, \aap\ {\bf 275}, L5 (1993).
\bibitem[Delbouille et al.(1973)]{Delbouille1973}
L. Delbouille, G. Roland, L. Neven, Atlas photometrique du spectre solaire de [lambda] 3000 a [lambda] 10000 (Liege: Universite de Liege, Institut d'Astrophysique) (1973)
\bibitem[Drawin(1969)]{Drawin1969}
H. W. Drawin, Z. Physik {\bf 225}, 483 (1969)
\bibitem[Fossati et al.(2009)]{Fossati2009}
L. Fossati, T. Ryabchikova, S. Bagnulo, E. Alecian, J. Grunhut, O. Kochukhov, G. Wade, \aap\ {\bf 503}, 945 (2009).
\bibitem[Giddings, 1981]{giddings81}
J. Giddings, Ph.D. Thesis, University of London (1981).
\bibitem[Gray(1977)]{gray77}
D. F. Gray, \apj\ {\bf 218}, 530 (1977).
\bibitem[Gustafsson et al., 2008]{2008A&A...486..951G}
B. Gustafsson, B. Edvardsson, K. Eriksson, U.G. Jorgensen, A. Nordlund, and B. Plez, \aap\ {\bf 486}, 951 (2008).
\bibitem[Holweger and Mueller(1974)]{HM74}
H. Holweger and E. A. Mueller, Solar Phys. {\bf 39}, 19 (1974).
\bibitem[Kochukhov, 2018]{Kochukhov_binmag}
O. Kochukhov, Astrophysics Source Code Library, record ascl:1805.015 (2018)
\bibitem[Kramida et al., 2019]{nist}
A. Kramida, Y. Ralchenko, J. Reader, NIST ASD Team, NIST Atomic Spectra Database (version 5.7.1). Gaithersburg MD, USA (2019).
\bibitem[Kurucz, 1993]{K1993}
R. Kurucz, ATLAS-9 model atmospheres. CD-ROM, Harvard-Smithsonian Center for Astrophysics (1993).
\bibitem[Kurucz, 2014]{K09}
R. Kurucz, Kurucz on-line database of observed and predicted atomic transitions, {\tt http://kurucz.harvard.edu/atoms/0700/}, (2014).
\bibitem[Kurucz et al., 1984]{KPNO1984}
R.~L. Kurucz, I, Furenlid, J. Brault, and L. Testerman, Solar Flux Atlas from 296 to 1300 nm Nat. Solar Obs., Sunspot, New Mexico  (1984).
\bibitem[Lemke and Venn(1995)]{Lemke95}
M. Lemke and K. A. Venn, \aap\ {\bf 309}, 558 (1996).
\bibitem[Lodders, 2021]{lodders21}
K. Lodders, Space Sci. Rev. {\bf 217}, id.44 (2021).
\bibitem[Lyubimkov et al.(2011)]{korotin_n1}
L.S. Lyubimkov, D.L. Lambert, S.A. Korotin, D.B. Poklad, T.M. Rachkovskaya,  S.I. Rostopchin, \mnras\ {\bf 410}, 1774 (2011).
\bibitem[Magg et al.(2022)]{Magg2022}
E. Magg, M. Bergemann, A. Serenelli, M. Bautista, B. Plez, U. Heiter, J. M. Gerber, H.-G. Ludwig, S. Basu, J. W. Ferguson, H. Carvajal Gallego, S. Gamrath, P. Palmeri, P. Quinet, \aap\ {\bf 661}, A140 (2022).
\bibitem[Mashonkina et al.(2020)]{mash_a20}
L. Mashonkina, T. Ryabchikova, S. Alexeeva, T. Sitnova, O. Zatsarinny, \mnras\ {\bf 499}, 3706 (2020).
\bibitem[Mashonkina(2024)]{mash_sc}
L. Mashonkina, \mnras\ {\bf 527}, 8234 (2024).
\bibitem[Meyer et al.(1997)]{Meyer1997}
D. M. Meyer, J. A. Cardelli, U. J. Sofia, \apj\ {\bf 490}, L103 (1997).
\bibitem[Neckel and Labs, 1984]{Neckel84}
H. Neckel and D. Labs, Sol. Phys. {\bf 90}, 205 (1984).
\bibitem[Netopil et al.(2022)]{NOC22}
M. Netopil, {\.I}. A. Oralhan, H. {\c{C}}akmak, R. Michel, Y. Karata{\c{s}}, \mnras\ {\bf 509}, 421 (2022).
\bibitem[Nieva and Przybilla(2012)]{Nieva2012}
M.-F. Nieva and N. Przybilla, \aap\ {\bf 539}, A143 (2012).
\bibitem[Pakhomov et al., 2019]{vald_hfs}
Yu. V. Pakhomov, T. A. Ryabchikova, N. E. Piskunov, Astron. Rep. {\bf 63}, 1010 (2019).
\bibitem[Przybilla and Butler(2001)]{Przybilla_n1}
N. Przybilla and K. Butler, \aap\ {\bf 379}, 955 (2001).
\bibitem[Przybilla et al.(2011)]{Przybilla2011}
N. Przybilla, M.-F. Nieva, K. Butler, Journal of Physics Conference Series, \textbf{328}, 012015 (2011).
\bibitem[Rentzsch-Holm(1996)]{n1_sun1996}
I. Rentzsch-Holm, \aap\ {\bf 305}, 275 (1996).
\bibitem[Romanovskaya et al.(2023)]{Roman2023}
A. Romanovskaya, T. Ryabchikova, Yu. Pakhomov, S. Korotin, T. Sitnova, \mnras\ {\bf 525},  3386 (2023).
\bibitem[Ryabchikova et al., 2015]{2015PhyS...90e4005R}
T. Ryabchikova, N. Piskunov, R. L. Kurucz, H. C. Stempels, U. Heiter, Y. Pakhomov, P. S. Barklem,  Phys. Scr., {\bf 90}, 054005 (2015).
\bibitem[Ryabchikova et al.(2016)]{Ryabchikova2016}
T. Ryabchikova, N. Piskunov, Yu. Pakhomov, V. Tsymbal, A. Titarenko, T. Sitnova, S. Alexeeva, L. Fossati and L. Mashonkina, \mnras\ {\bf 456}, 1221 (2016).
\bibitem[Ryabchikova et al.(2022)]{Ryabchikova_CN}
T. Ryabchikova, N. Piskunov, Y. Pakhomov, Atoms {\bf 10}, 103 (2022).
\bibitem[Seaton(1962)]{seaton62}
M. J. Seaton, in Atomic and Molecular Processes (New York: Academic Press) (1962).
\bibitem[Seaton, 1987]{seaton87}
M. J. Seaton, Journal of Physics B: Atomic Molecular Physics {\bf 20}, 6363 (1987).
\bibitem[Shulyak et al., 2004]{llmodels}
D. Shulyak, V. Tsymbal, T. Ryabchikova, C. St{\"u}tz, W. W. Weiss, \aap\ {\bf 428}, 993 (2004).
\bibitem[Sitnova and Mashonkina(2018)]{o_sun2018}
T. M. Sitnova and L. I. Mashonkina, Astron. Lett. {\bf 44}, 411 (2018).
\bibitem[Sitnova et al.(2022)]{zn2022}
T. M. Sitnova, S. A. Yakovleva, A. K. Belyaev, L. I. Mashonkina, \mnras\ {\bf 515}, 1510 (2022).
\bibitem[Tachiev and Froese Fischer(2002)]{n1_nist}
G. I. Tachiev and C. Froese Fischer, \aap\ {\bf 385}, 716 (2002).
\bibitem[Takeda(1992)]{takeda1992}
Y. Takeda, Publ. Astron. Soc. Japan {\bf 44}, 649 (1992).
\bibitem[Takeda et al.(2007)]{takeda2007}
Y. Takeda, S. Kawanomoto, N. Ohishi, Publ. Astron. Soc. Japan {\bf 59}, 245 (2007).
\bibitem[Takeda et al.(2018)]{takeda2018}
Y. Takeda, S. Kawanomoto, N. Ohishi, D.-II Kang, B.-C. Lee, K.-M. Kim, I. Han, Publ. Astron. Soc. Japan {\bf 70}, 91 (2018).
\bibitem[Tsymbal et al., 2019]{detail12}
V. Tsymbal, T. Ryabchikova, T. Sitnova, in Kudryavtsev D.O., Romanyuk I.I., Yakunin I.A., eds, Astron. Soc. Pacific Conf. Ser. {\bf 518}. Physics of Magnetic stars, San Francisco: Astronomical Society of the Pacific, 247 (2019)
\bibitem[van Regemorter, 1962]{Reg1962}
H. van Regemorter, Astrophys. J., \textbf{136}, 906 (1962).
\bibitem[Venn and Lambert(1990)]{Venn90}
K.A. Venn and D.L. Lambert, \apj\ {\bf 363}, 234 (1990).
\bibitem[Wallace et al.(2011)]{Sun_2011}
L. Wallace, K.~H. Hinkle, W.~C. Livingston, S.~P. Davis, Astrophys. J. Suppl. Ser. {\bf 195}, 6 (2011).
\bibitem[Wang et al.(2014)]{Wang_n1}
Y. Wang, O. Zatsarinny, K. Bartschat, Phys. Rev. {\bf A89}, 062714 (2014).
\end{thebibliography}
\end{document}